\documentclass[intlimits,twoside,a4paper]{article}

\usepackage[cp1251]{inputenc}

\usepackage[]{cmpj3}

\usepackage{bm}

\newcommand{\bea}{\begin{eqnarray}}
\newcommand{\eea}{\end{eqnarray}}
\newcommand{\be}{\begin{equation}}
\newcommand{\ee}{\end{equation}}
\newcommand{\eps}{\varepsilon}

\issue{2020}{23}{3}{33704}
\doinumber{10.5488/CMP.23.33704}
\title[Effects of diagonal strains and H-bond geometry in antiferroelectric 
	squaric acid crystals]%
{Effects of diagonal strains and H-bond geometry in antiferroelectric 
	squaric acid crystals%
}
\author{A.P. Moina}
\address{
Institute for Condensed Matter Physics of the National
Academy of Sciences of Ukraine,\\ 1 Svientsitskii St., 79011 Lviv,
Ukraine
}
\date{Received December  14, 2019, in final form April 26, 2020}

\begin{document}

\maketitle

\begin{abstract}
The proton ordering model of the phase transition and physical properties of antiferroelectric crystals of squaric acid is modified by taking into account the influence of diagonal lattice strains and of the local geometry of hydrogen bonds, namely of the distance $\delta$ between the H-sites on a bond. Thermal expansion, the spontaneous strain $\eps_1-\eps_3$, and specific heat of squaric acid are well described by the proposed model. However, a consistent description of hydrostatic pressure influence on the transition temperature is possible only with further modifications of the model.

\end{abstract}

\section{Introduction}

The crystals of squaric acid, H$_2$C$_4$O$_4$ (3,4-dihydroxy-3-cyclobutene-1,2-dione) are an epitome of two-dimensional antiferroelectrics. The hydrogen bonded C$_4$O$_4$ groups form  planes parallel to $ac$ and stacked along the $b$-axis. Below the transition  at 373~K, a spontaneous polarization arises in these planes, with the neighbouring planes polarized in the opposite directions. The crystal symmetry changes from centrosymmetric tetragonal, $I4/m$, to monoclinic, $P2_1/m$, and spontaneous symmetry-changing strains $\eps_1-\eps_3$ (orthorhombic) and $\eps_5$ (monoclinic), both of $B_g$ symmetry, arise \cite{semmingsen:77,semmingsen:95,hollander:77}. The local symmetry of the H$_2$C$_4$O$_{4}$ groups is C$_{1h}$ below and above the antiferroelectric phase transition \cite{moritomo:91}.

Elastic and thermoelastic properties of squaric acid are remarkably anisotropic. Compressibility and thermal expansion \cite{ehses:81} are much higher in a direction perpendicular to the planes of hydrogen bonds than within the planes. The symmetry-changing strains $\eps_1-\eps_3$ and $\eps_5$ are confined to the $ac$ plane. The anomalous parts of the diagonal strains $\eps_1$, $\eps_3$ and of
$\eps_2$, caused by electrostriction, have different signs. 
%Being caused by electrostriction, anomalous parts of the diagonal strains $\eps_1$, $\eps_3$ and of $\eps_2$ have different signs~\cite{ehses:81}.

There is also experimental evidence for non-equivalence of hydrogen bonds going along  two perpendicular directions (e.g. \cite{klymachyov:97,semmingsen:95}). The difference between degrees of proton ordering on these bonds is about 2\% at $T_{\textrm N}-13$~K and $T_{\textrm N}-21$~K \cite{semmingsen:95}. The O--H and H-site distances are also found to be slightly different for the perpendicular bonds.

Hydrostatic pressure rapidly decreases the antiferroelectric transition temperature with the slope of about 11~K/kbar \cite{samara:79,yasuda:78,moritomo:91} at pressures below about 25 kbar. At higher pressures, this dependence deviates from linearity, and around 28~kbar, the transition temperature rapidly falls to zero: a quantum paraelectric state is induced \cite{moritomo:91}.  At further increase of pressure at a constant temperature (at least for temperatures between 100 and 300 K), another phase boundary is detected \cite{moritomo:91}, at which the local symmetry of the H$_2$C$_4$O$_{4}$ groups changes from  C$_{1h}$ to C$_{4h}$.

Theoretical description of the antiferroelectric transition in squaric acid is usually based on some versions of the proton ordering model, either two-dimensional, invoking four-particle correlations between protons within the planes \cite{matsushita:80,matsushita:81,matsushita:82,chaudhuri:90}, or one-dimensional, where either non-interacting \cite{maier:82} or coupled \cite{deininghaus:81} perpendicular pseudospin chains are considered. The four-particle model can be reduced to the model of interacting one-dimensional chains by the proper choice of the  model parameters \cite{maier:82}. The four-particle Hamiltonians are basically identical to those for NH$_4$H$_2$PO$_4$ crystals, antiferroelectrics of the KH$_2$PO$_4$ family. 

Deformational effects in squaric acid were first addressed in \cite{matsushita:80,matsushita:81,matsushita:82}, where the coupling between spins and spontaneous lattice distortion were included into the model.  
In \cite{chaudhuri:90} the proton-phonon coupling was added, and hydrostatic pressure effects on the phase transition temperature and dielectric permittivity of pure and deuterated squaric acid crystals were described by assuming the model parameters to be pressure dependent and by performing a new fitting procedure for each considered value of pressure.

In \cite{ishizuka:11,vijigiri:20}, the observed in \cite{moritomo:91} $p-T$ phase diagram comprising the antiferroelectric phase with the C$_4$O$_4$ groups having the local symmetry C$_{1h}$, the paraelectric phase (C$_{1h}$), and high-pressure paraelectric phase (C$_{4h}$), was qualitatively described, using different versions of proton ordering model. The intermediate paraelectric phase (C$_{1h}$) was found to be a quantum liquid-like state.

Since the antiferroelectric phase transition in squaric acid is usually attributed to proton ordering, which triggers displacements of heavy ions and rearrangement of electronic density, it is expected that just like in the KH$_2$PO$_4$ family crystals, in the squaric acid the pressure-induced changes in the geometry of the hydrogen bonds should  play an important role in the pressure influence on the phase transition. The distance $\delta$ between the two equilibrium positions of a proton on a hydrogen bond was found to be the most crucial geometrical parameter here \cite{mcmahon:90,stasyuk:99}.
For the KH$_2$PO$_4$ family crystals, having a three-dimensional network of hydrogen bonds, there exists a universal linear dependence of the transition temperatures $T_c$ on the value of the distance $\delta$ at the transition, with $T_c$ and $\delta$ being varied by external hydrostatic pressure, uniaxial stress $p=-\sigma_3$, and by isomorphic substitution of heavy ions~\cite{mcmahon:90,stasyuk:99}. For the squaric acid, which H-bond network is two-dimensional,  the dependence $T_c(\delta)$ under hydrostatic pressure is similar but with a different slope. 
 
None of the above mentioned  earlier theories for the squaric acid crystals explicitly considers the role of the geometrical parameters of hydrogen bonds in the pressure effects on the phase transition in squaric acid. None of them includes into consideration the thermal expansion of the crystal  either.

Thus, similarly to how it was done for Rochelle salt \cite{moina:11}, we intend to develop a unified deformable model for squaric acid that can describe the effects associated with the diagonal lattice strains: thermal expansion and influence of external hydrostatic pressure. We shall also include the dependence of the interaction constants on the H-site distance $\delta$ into the model. 

An important question arises whether tunneling of protons on hydrogen bonds, also known to be essentially dependent on their geometry, should be taken into account in the model. We believe that for the reasons described below, at temperatures and pressures considered in the present paper  it will suffice to use an Ising-type model without tunneling.

The calculations performed within the framework of the proton-lattice model \cite{chaudhuri:90} using the random phase approximation found tunneling to be small in squaric acid, even when it is increased by hydrostatic pressure, as it is usually assumed. Furthermore, when the cluster approximation for the short-range interactions is used instead of the  mean-field type approximations (MFA), tunneling becomes even less essential. Thus, for the KH$_2$PO$_4$ type systems, tunneling is effectively renormalized by short-range four-particle correlations between protons \cite{blinc:87}, reducing its effective value down to tenths of that used by the MFA calculations. Tunneling is expected to be significant at very low temperatures and at high pressures, where quantum fluctuations suppress the macroscopic ordering, leading to the onset of quantum paraelectricity. We, on the other hand, restrict our consideration to moderate pressures (below 15~kbar) and higher temperatures (above 150~K), where the $T_c(p)$ dependence remains linear. We think it safe to assume that in our calculations tunneling can be neglected.

\section{The model}
\begin{figure}[htb]
\centerline{\includegraphics[height=0.65\textwidth]{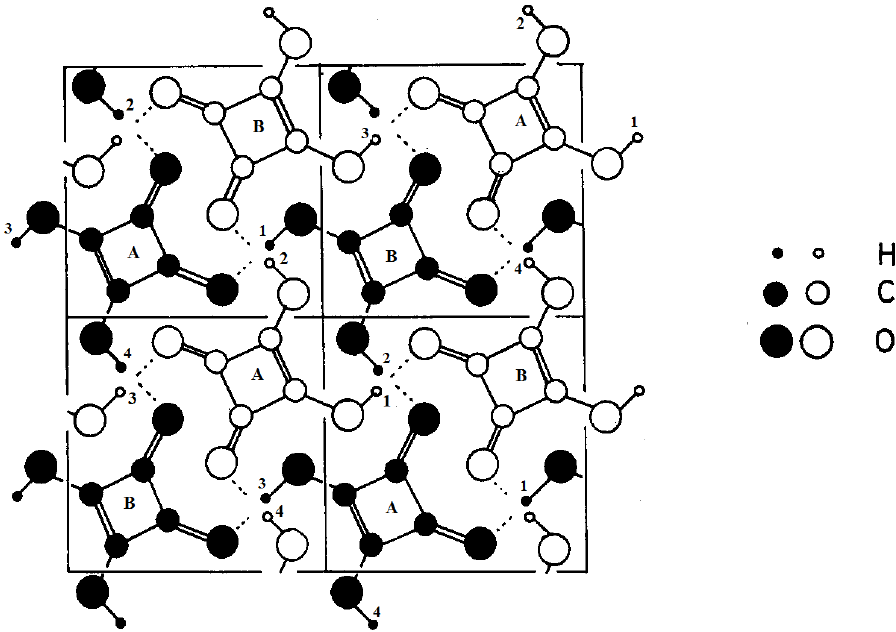}}
		\caption{Crystal structure of squaric acid as viewed along the $b$ axis. Figure is taken from \cite{semmingsen:74,moritomo:90}. Two adjacent layers are shown, with black and open circles each. The A and B type C$_4$O$_4$ groups are  indicated (see text for explanation), and the hydrogen bonds are numbered.} \label{sqa-structure}
\end{figure}

There are two formula units in the low-temperature phase unit cell of squaric acid. In our model, the unit cell consists of two C$_4$O$_4$ groups and four hydrogen atoms ($f=1,2,3,4$, see figure~\ref{sqa-structure}) attached to one of them (the A type group). All hydrogens around the B type groups are considered to belong to the A type groups, with which the B groups are hydrogen bonded. Note that the two C$_4$O$_4$ groups of each unit cell belong to
different neighboring layers. The center of each hydrogen bond lies exactly above   the center of the hydrogen bond in the layer below it (as seen along the $b$ axis). The bonds around each A type group are numbered counterclockwise.

As it is usual in the proton ordering models, we consider interactions between protons leading to an ordering in their system. Motion of protons in double-well potentials is described by pseudospins, whose two eigenvalues $\sigma=\pm 1$ are assigned to two equilibrium positions of the proton.
We take into account the presence of the diagonal components of the lattice strain tensor $\eps_1$, $\eps_2$, and $\eps_3$ that are induced via thermal expansion or by application of external hydrostatic pressure.

The system Hamiltonian in the case of squaric acid
\begin{equation}\label{sqa-Ham}
H=U_{\textrm{seed}}+H_{\textrm{long}}^{\textrm{intra}}+H_{\textrm {long}}^{\textrm{inter}}+H_{\textrm{short}}
\end{equation}
includes ferroelectric intralayer long-range interactions $H_{\textrm{long}}^{\textrm {intra}}$, ensuring ferroelectric 
ordering within each separate layer, antiferroelectric interlayer $H_{\textrm{ long}}^{\textrm{inter}}$ responsible for alternation of polarizations in the stacked layers, the short-range configurational interactions between protons $H_{\textrm{short}}$, and the so-called ``seed'' energy
\begin{equation}\label{key2}
U_{\textrm{seed}}=vN\left[\frac 12\sum_{ij=1}^3c_{ij}^{(0)}\eps_i\eps_j-\sum_{ij=1}^3c_{ij}^{(0)}\alpha_i^{(0)}(T-T_i^0)\eps_j\right],
\end{equation}
containing elastic and thermal expansion contributions associated with uniform lattice strains; $c_{ij}^{(0)}$ are the corresponding ``seed'' elastic constants, whereas $\alpha_i^{(0)}$ are the ``seed'' thermal expansion coefficients. $T_i^0$ determines the reference point of the thermal expansion of the crystal, which can be chosen arbitrarily. $v$ is the unit cell volume, and $N$ is the number of the unit cells in the crystal. Such a form of (\ref{key2}) later on yields  standard expressions for the strains of a stressed thermally expanding solid [see equation~(\ref{eps-anom})].

\begin{figure}[!t]
	\centerline{\includegraphics[height=0.2\textwidth]{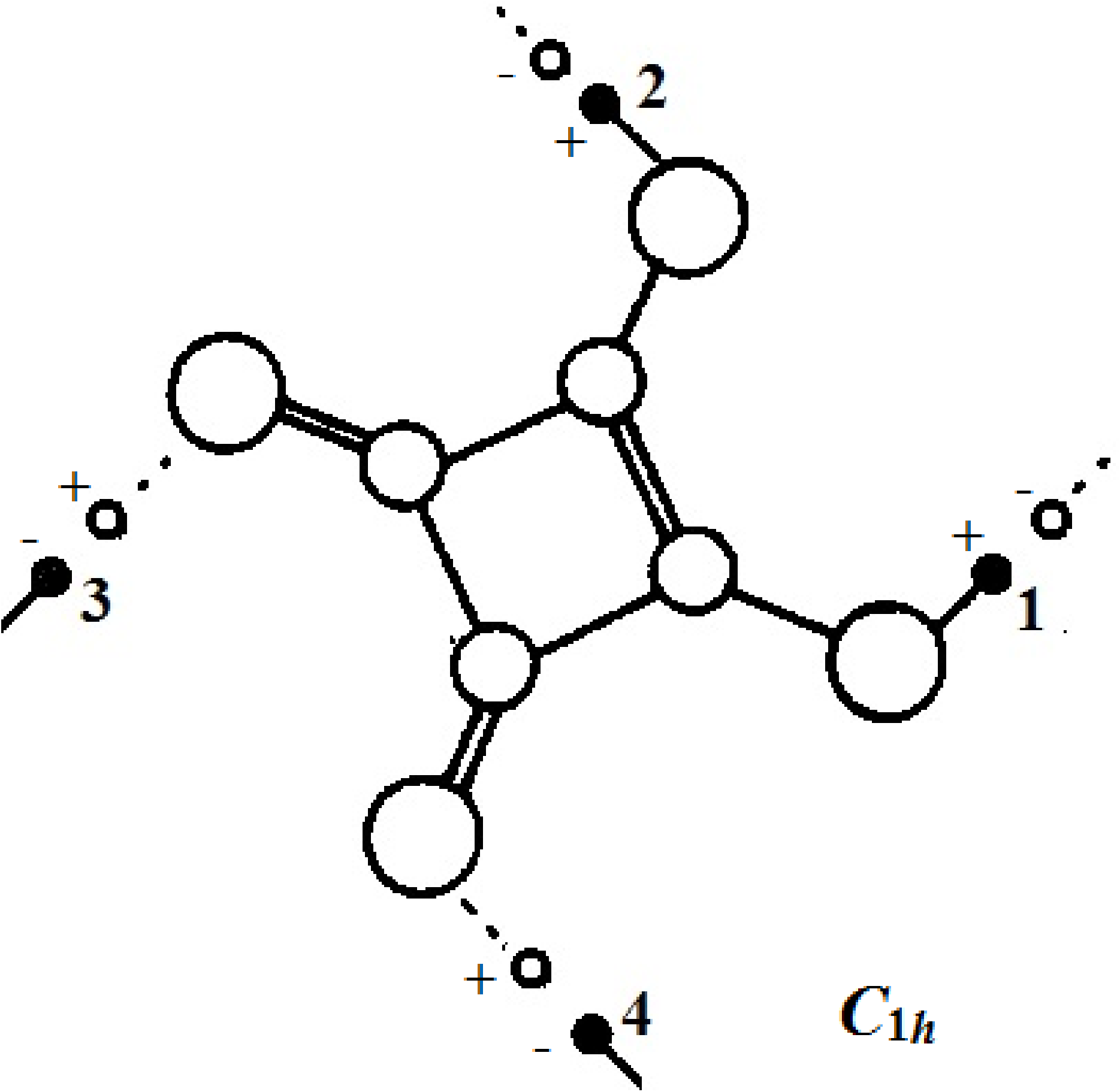}$\qquad$
		\includegraphics[height=0.2\textwidth]{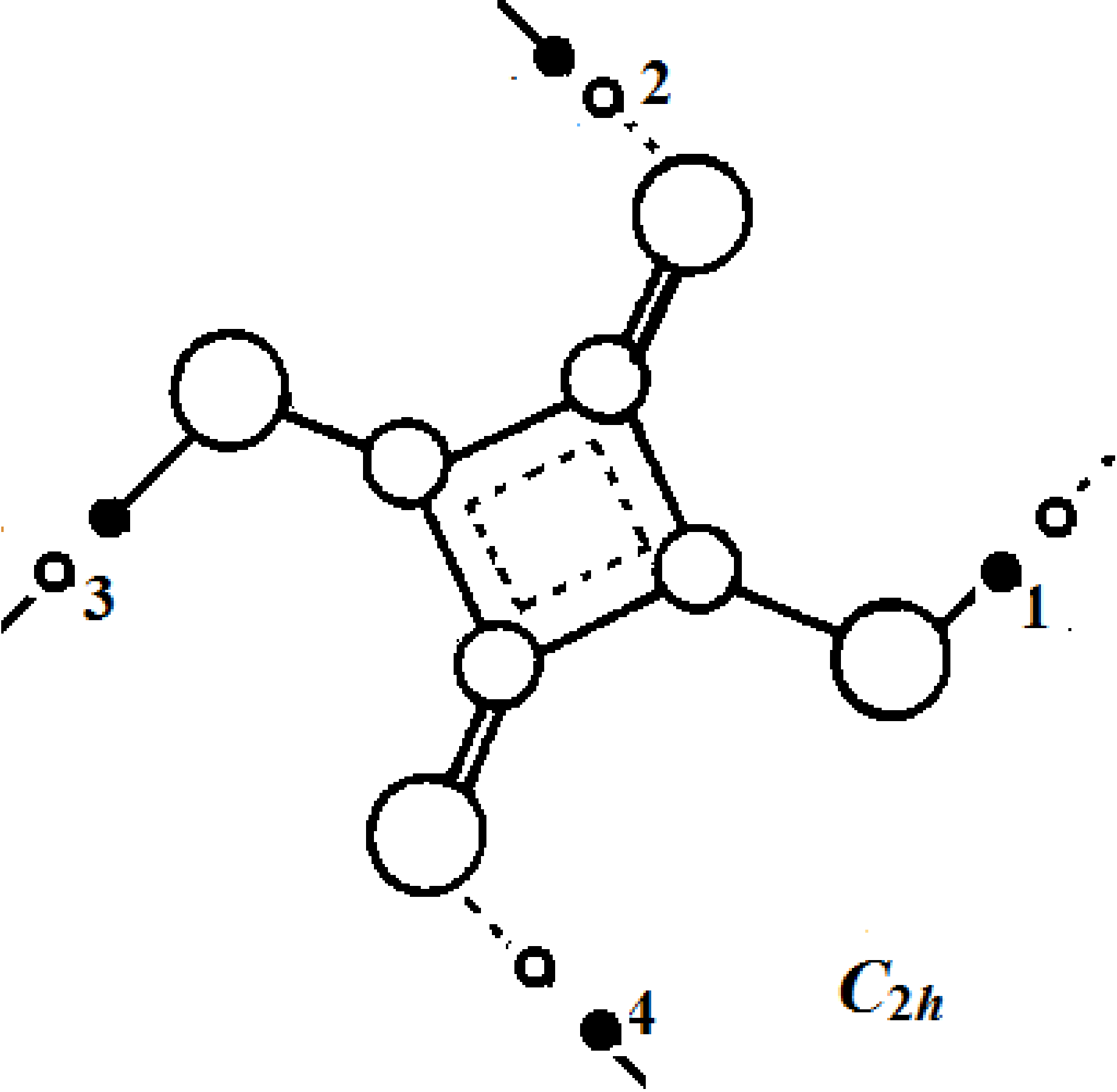}$\qquad$
		\includegraphics[height=0.2\textwidth]{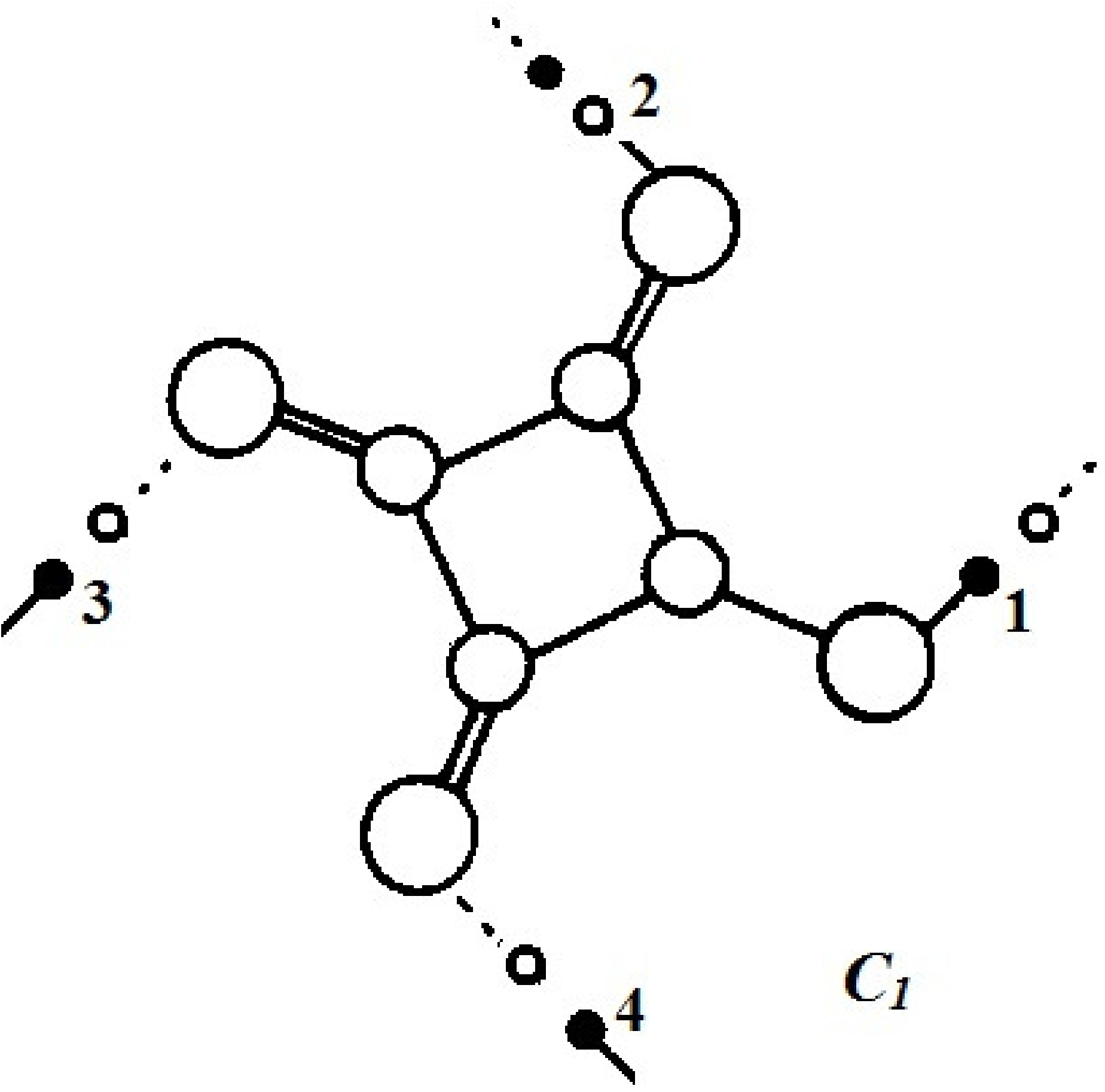}$\qquad$
		\includegraphics[height=0.2\textwidth]{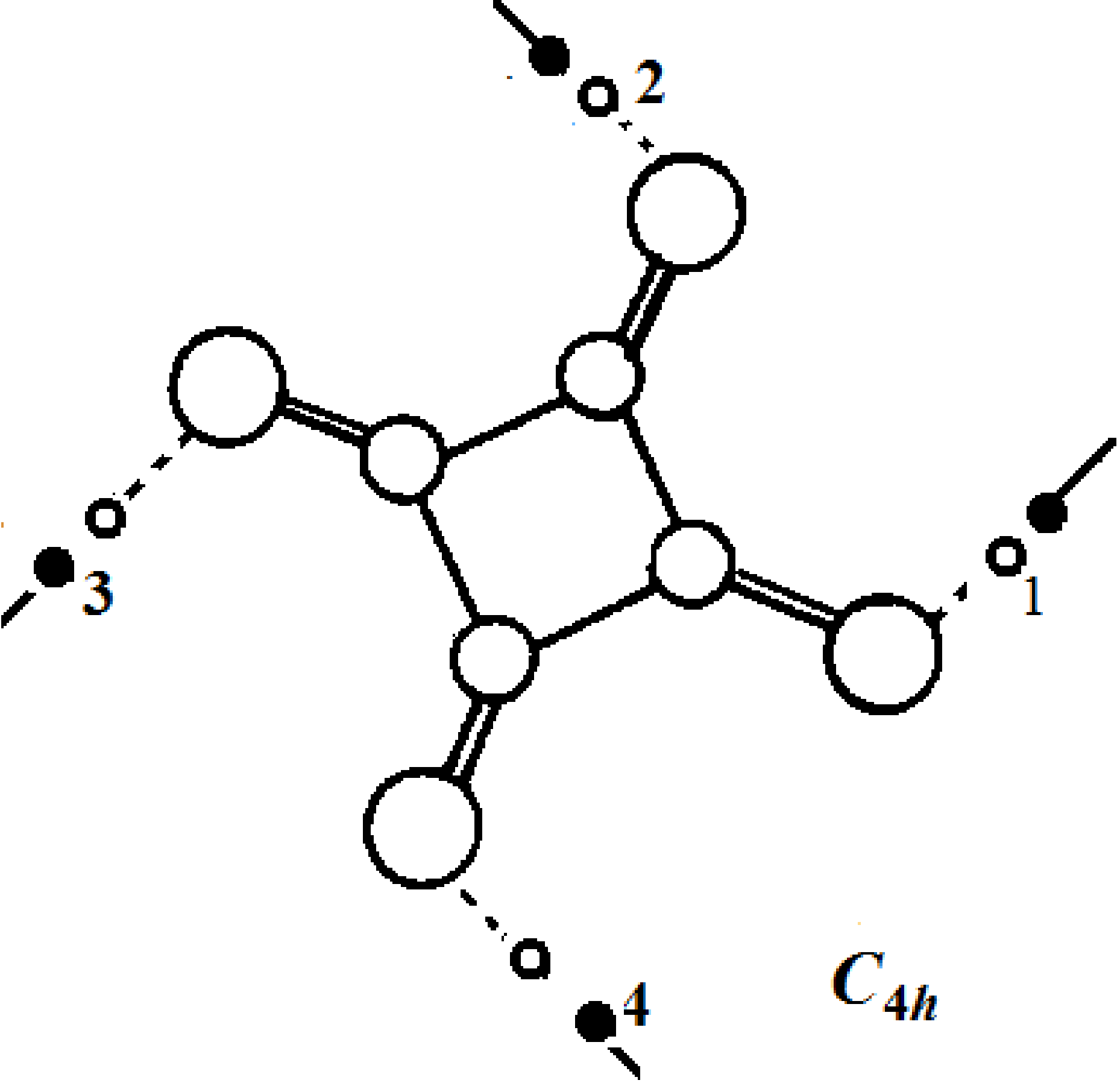}}
	\caption{Lateral, diagonal, single and double ionized proton configurations around
	an A type C$_4$O$_4$ group in squaric acid. The hydrogen bonds $f=1,2,3,4$
	are numbered. Two equilibrium positions of each proton are shown, and the
	signs $s_f=\pm 1$ of the eigenvalues of the $\sigma_{yqf}$ operators are
	indicated. Here $y$ stands for the layer index, $q$ is the index of the A
	type C$_4$O$_4$ group, and $f$ is the bond index.} \label{configurations_fig}
\end{figure}

The short-range Hamiltonian $H_{\textrm{short}}$ describes the four-particle confirational correlations between protons sitting around each C$_4$O$_4$ group. 
Similarly to how it is done for NH$_4$H$_2$PO$_4$, the antiferroelectrics of the KH$_2$PO$4$ type family, it is assumed that the energy of four lateral configurations $\eps_a$ is the lowest of all, where two protons are in positions close to the adjacent oxygens of the C$_4$O$_4$ group, whereas two other protons are closer to the neighboring C$_4$O$_4$ groups (see figure~\ref{configurations_fig}). The next level is two diagonal configurations with the energy $\eps_s$, where the protons are close to the opposite oxygens of the C$_4$O$_4$ group. Then, there are eight single-ionized configurations with three protons or only one close proton, having the energy $\eps_1$, and two double-ionized configurations ($\eps_0$) with four or no protons at all close to the given C$_4$O$_4$ group. It is believed that $\eps_a<\eps_s\ll\eps_1\ll\eps_0$.

If two protons are in the most energetically favorable lateral configurations, the C$_4$O$_4$ groups are isosceles trapezoids (point group C$_{1h}$), although very close to squares (C$_{4h}$). It is believed that this local distortion is caused by two single and two double alternating covalent bonds connecting the four oxygens to the carbons, and by formation of the double C--C bond within the C$_4$O$_4$ skeleton, as shown in figures~\ref{sqa-structure} and \ref{configurations_fig}. Double bonds are shorter than single ones between analogous atoms. Since the origin of the skeleton distortion is the chemical bonding with the local proton configuration, rather than macroscopic uniform lattice strains, all four lateral configurations still have the same energy of  short-range interactions, no matter what their orientation relatively to the crystallographic axes is (see table~\ref{configurations_table}). 
The same holds for the diagonal (point group C$_{2h}$), single-ionized 
(point group C$_{1}$), and double-ionized groups (point group D$_{2h}$). It means that no splitting of  short-range energy levels by the macroscopic spontaneous strain takes place, in contrast to what was assumed in earlier theories for squaric acid \cite{matsushita:80} or for KH$_2$PO$_4$ type crystals \cite{stasyuk:00,stasyuk:01}.

\renewcommand{\tabcolsep}{3.0pt}
\begin{table}[!t]
	\caption{Proton configurations and their energies; $\eps_a<\eps_s\ll\eps_1\ll\eps_0$. }
	\label{configurations_table}
	\begin{center}
		\small
		\begin{tabular}{c|c|c|c}
			\hline
			$i$ &   &  $s_1s_2s_3s_4$ &${\cal E}_i$ \\
			\hline
			1 & \includegraphics[height=1cm,width=1cm]{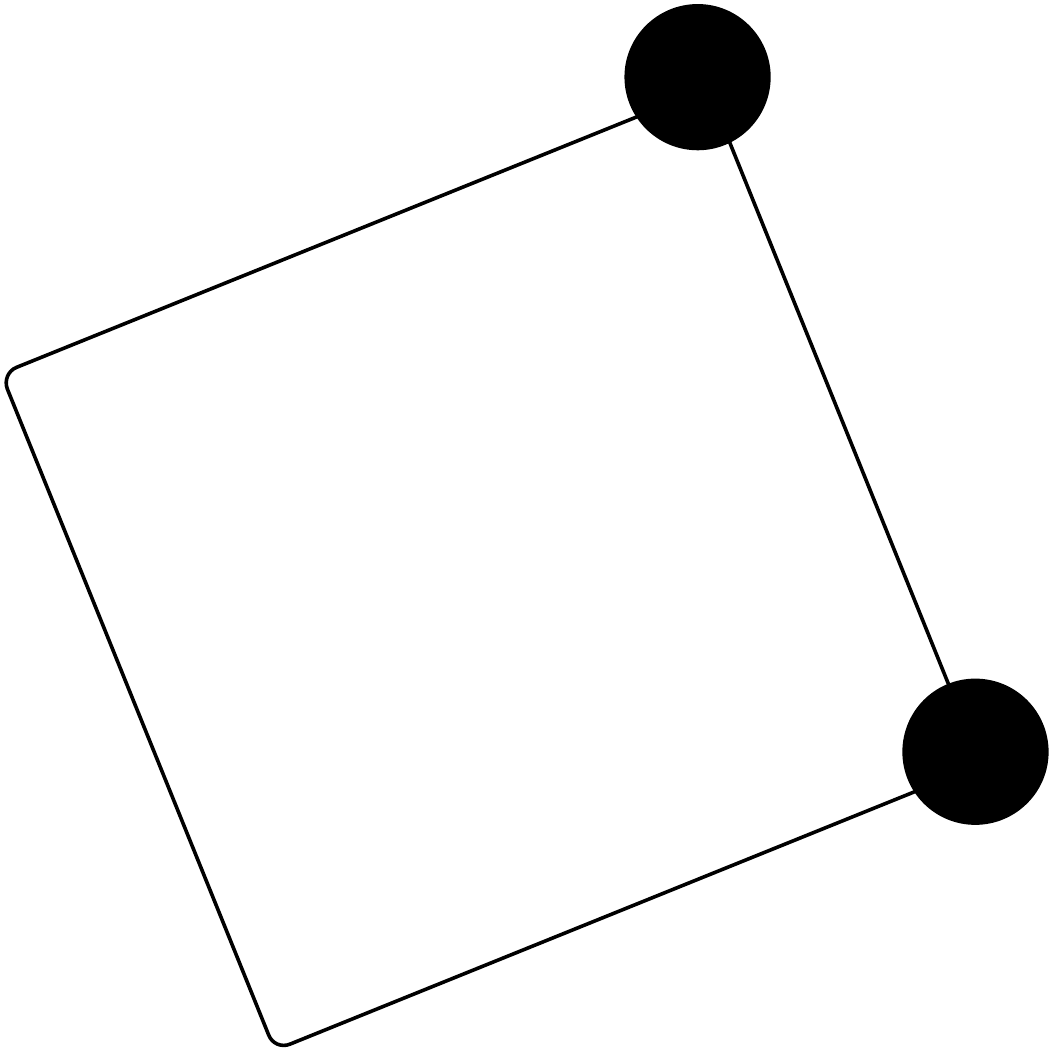}  &   $++--$ & $\eps_a$ \strut \\
			2 &  \includegraphics[height=1cm,width=1cm]{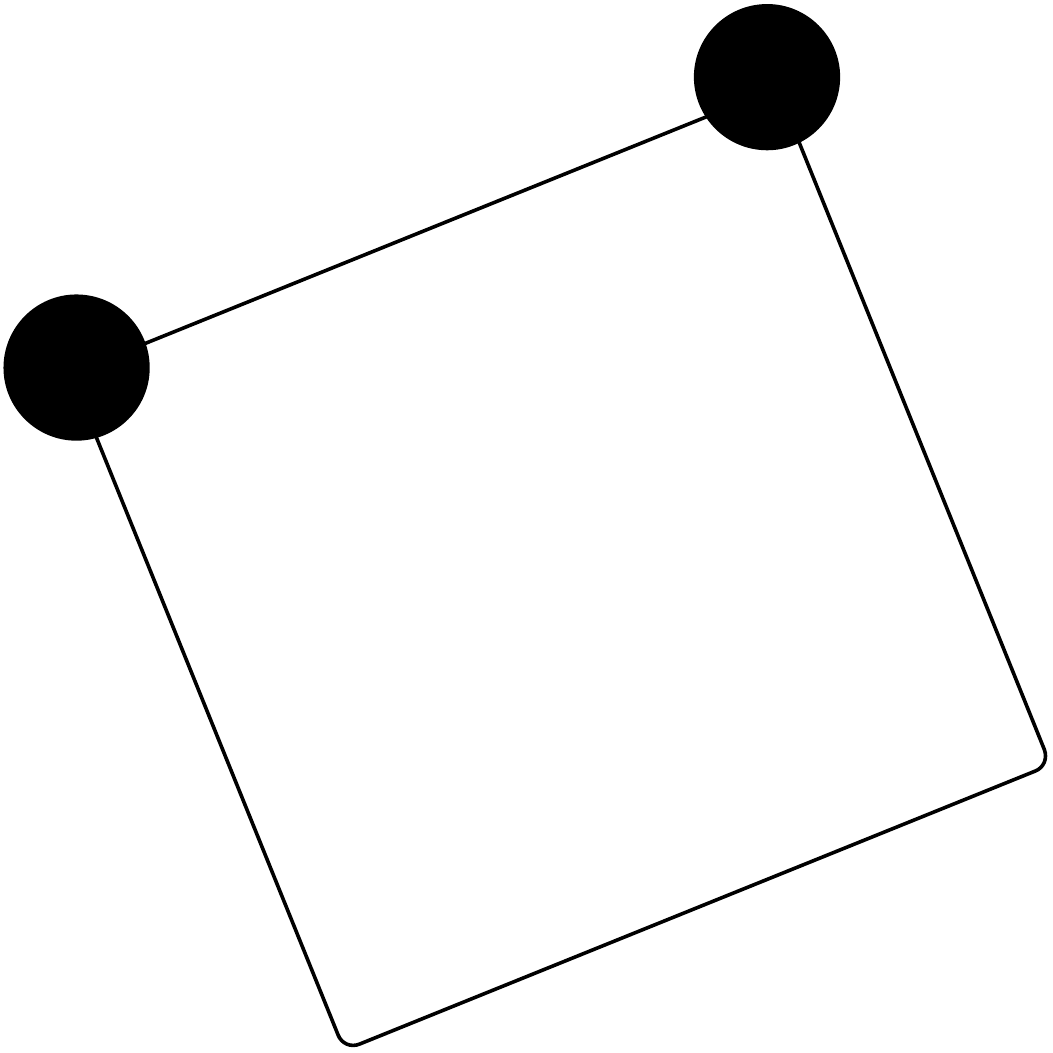} &   $-++-$  \\
			3 & \includegraphics[height=1cm,width=1cm]{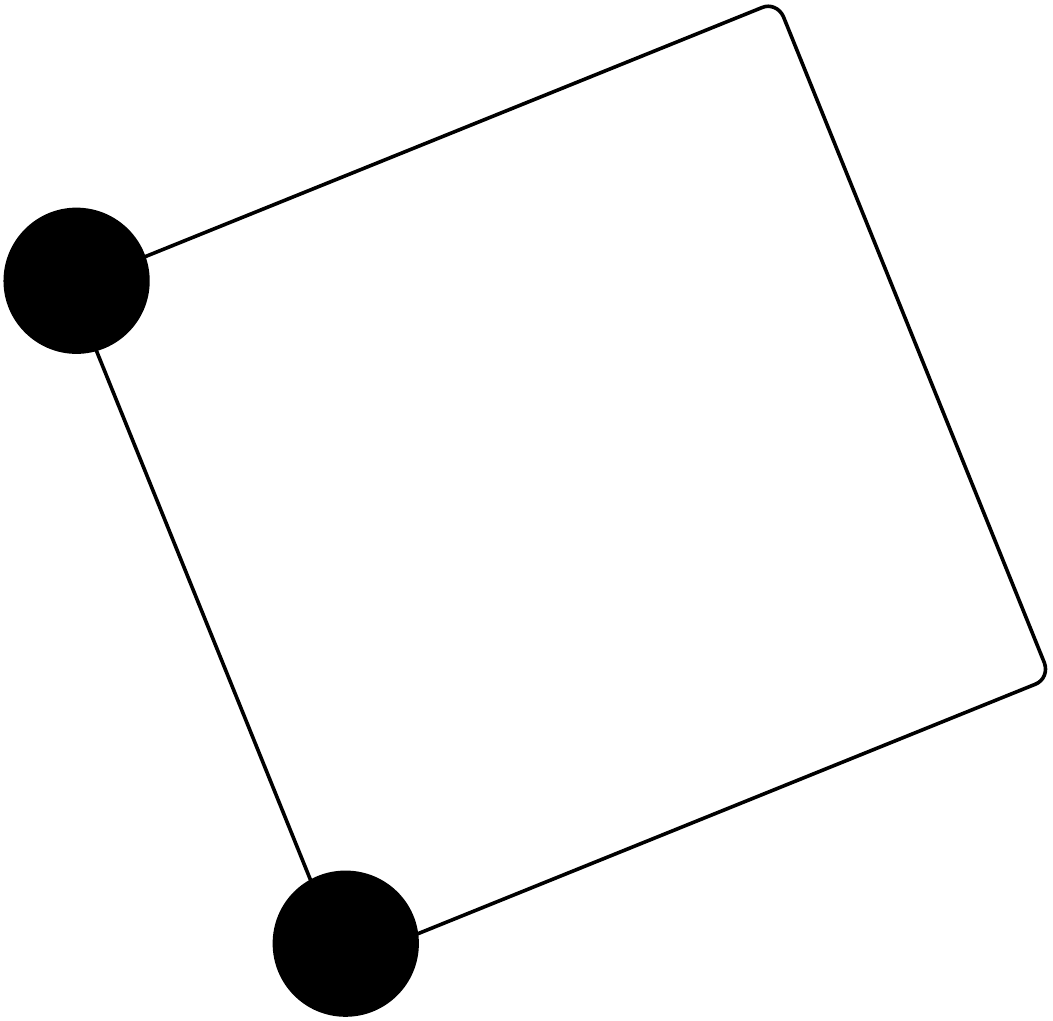}  &   $--++$&   \\
			4 &\includegraphics[height=1cm,width=1cm]{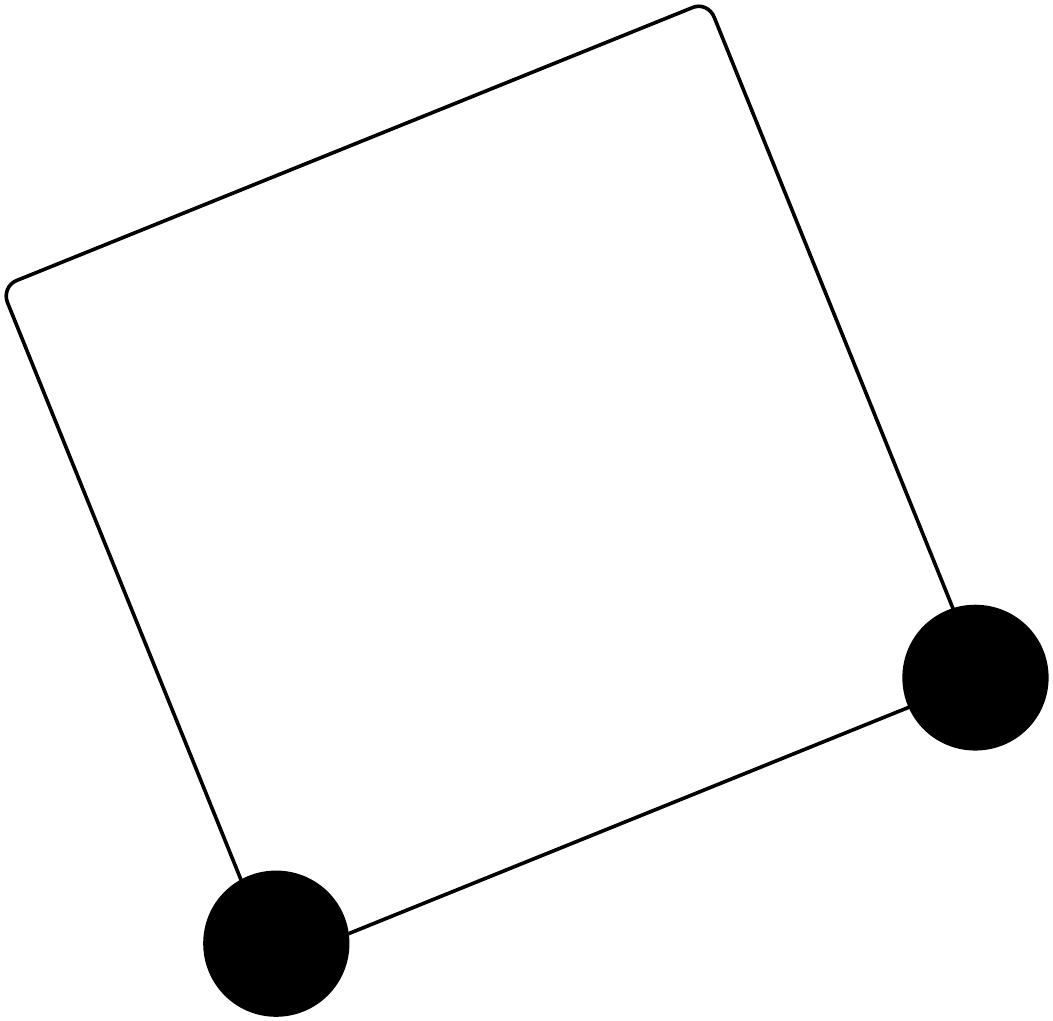} &  $+--+$  \\
			\hline
			5 & \includegraphics[height=1cm,width=1cm]{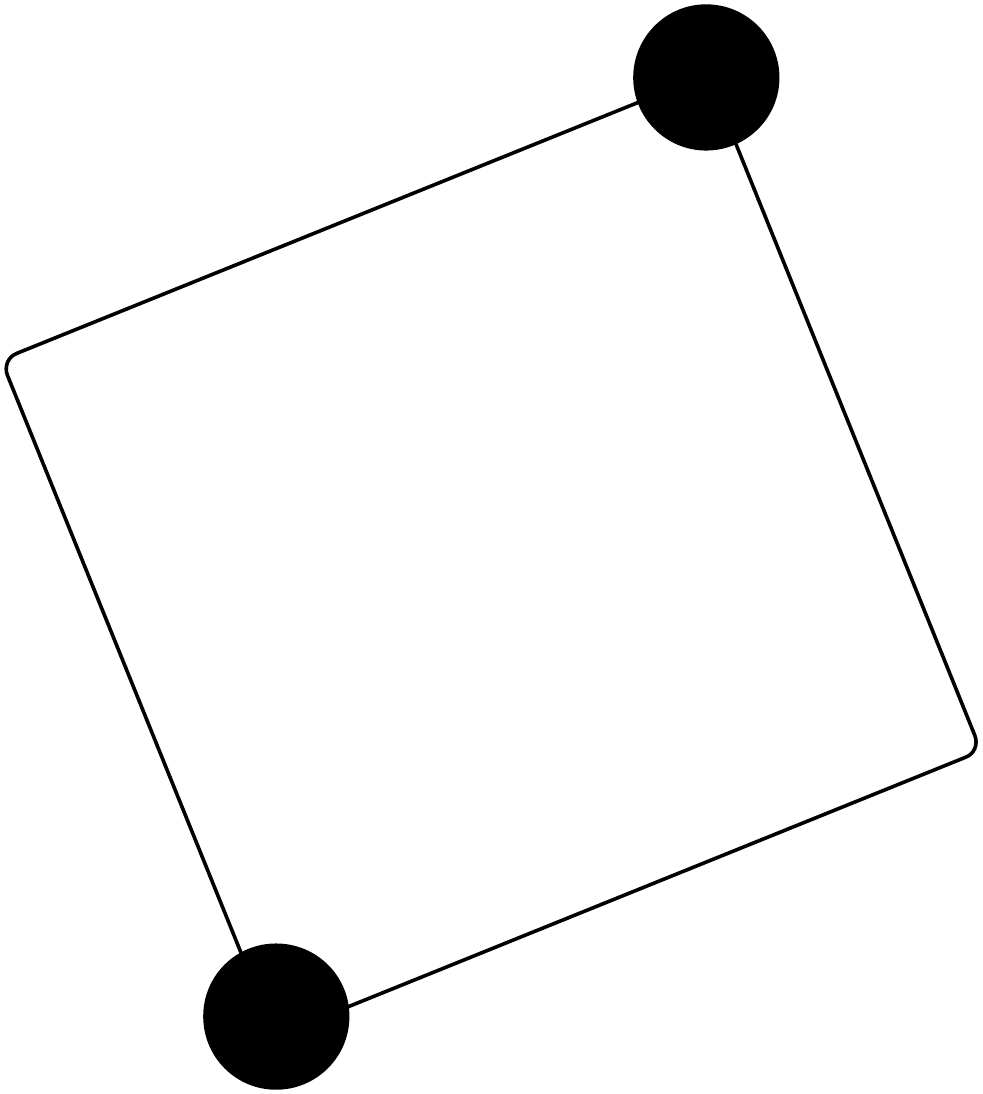} &  $-+-+$ & $\eps_s$ \\
			6 & \includegraphics[height=1cm,width=1cm]{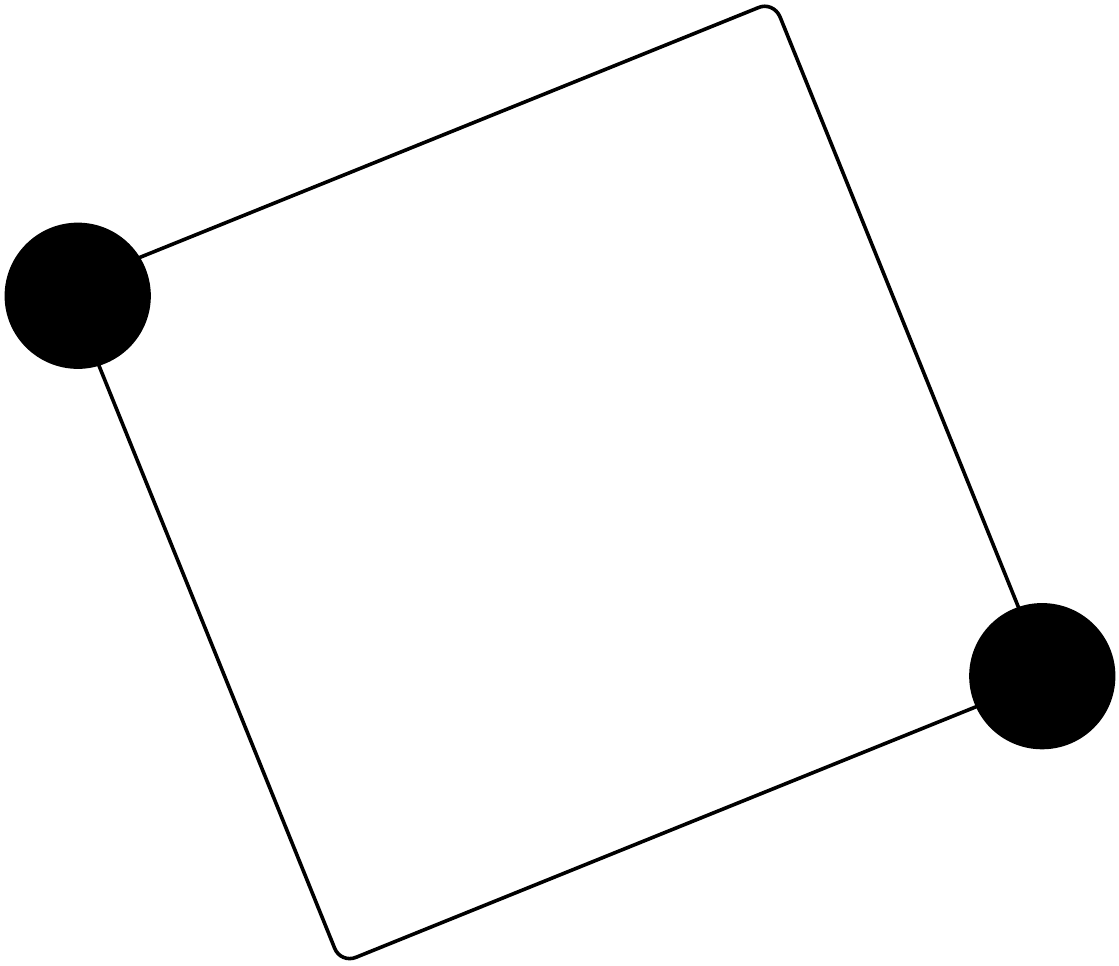} &  $+-+-$ & \\
			\hline
			15 & \includegraphics[height=1cm,width=1cm]{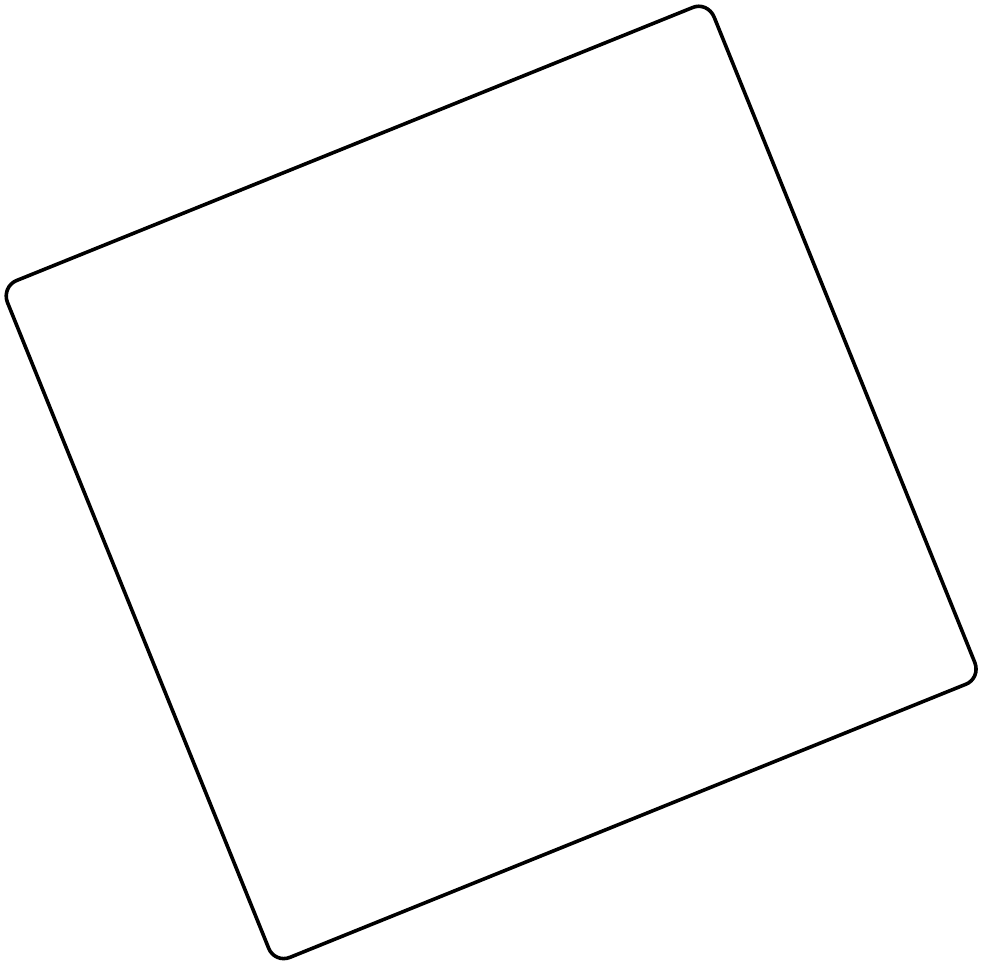}  &  $----$ & $\eps_0$  \\
			16 & \includegraphics[height=1cm,width=1cm]{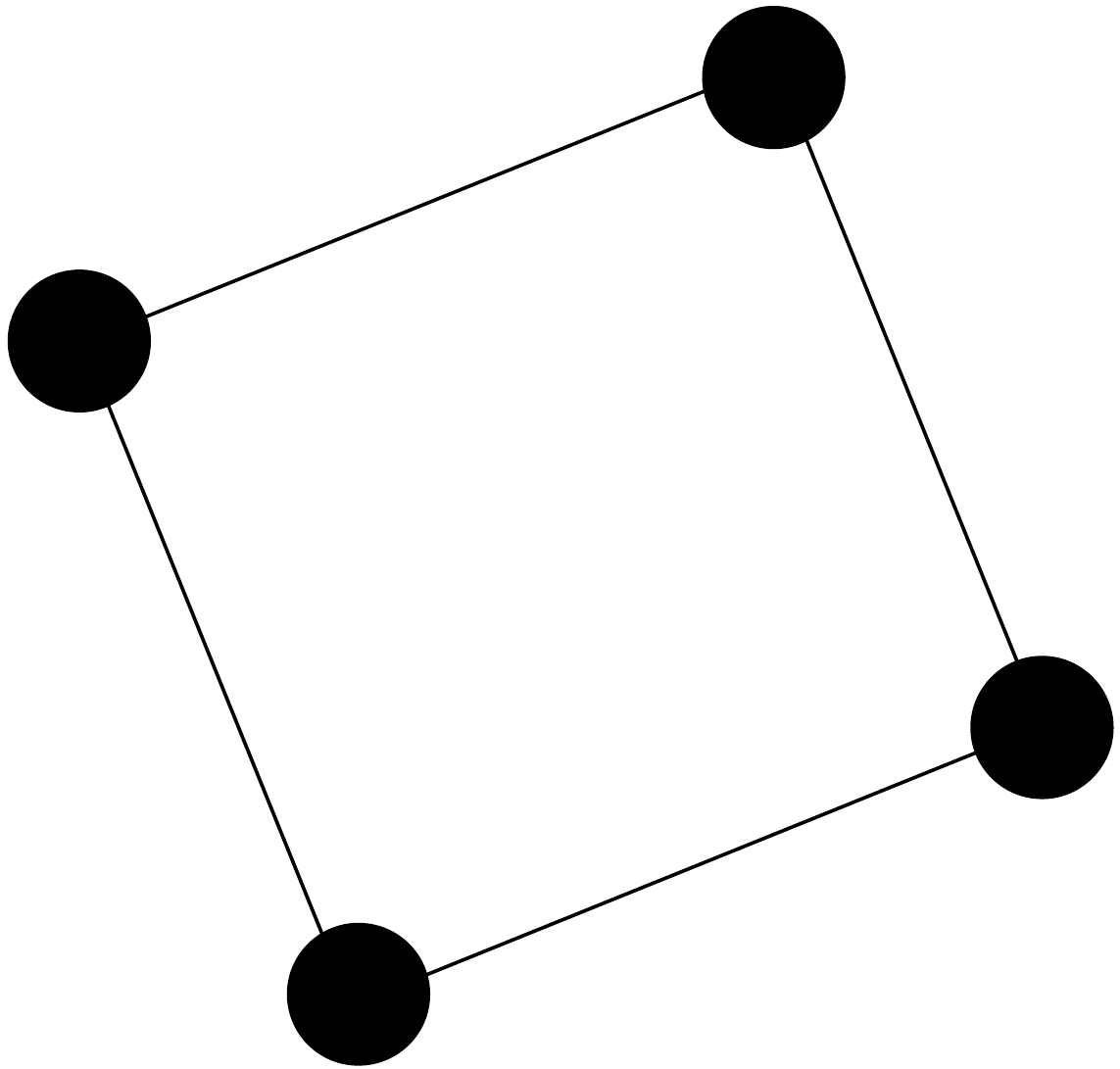} &  $++++$& \\
			\hline
		\end{tabular}\hspace{5ex}
		\begin{tabular}{c|c|c|c}
			\hline
			$i$ & & $s_1s_2s_3s_4$ &${\cal E}_i$ \\
			\hline
			7 & \includegraphics[height=1cm,width=1cm]{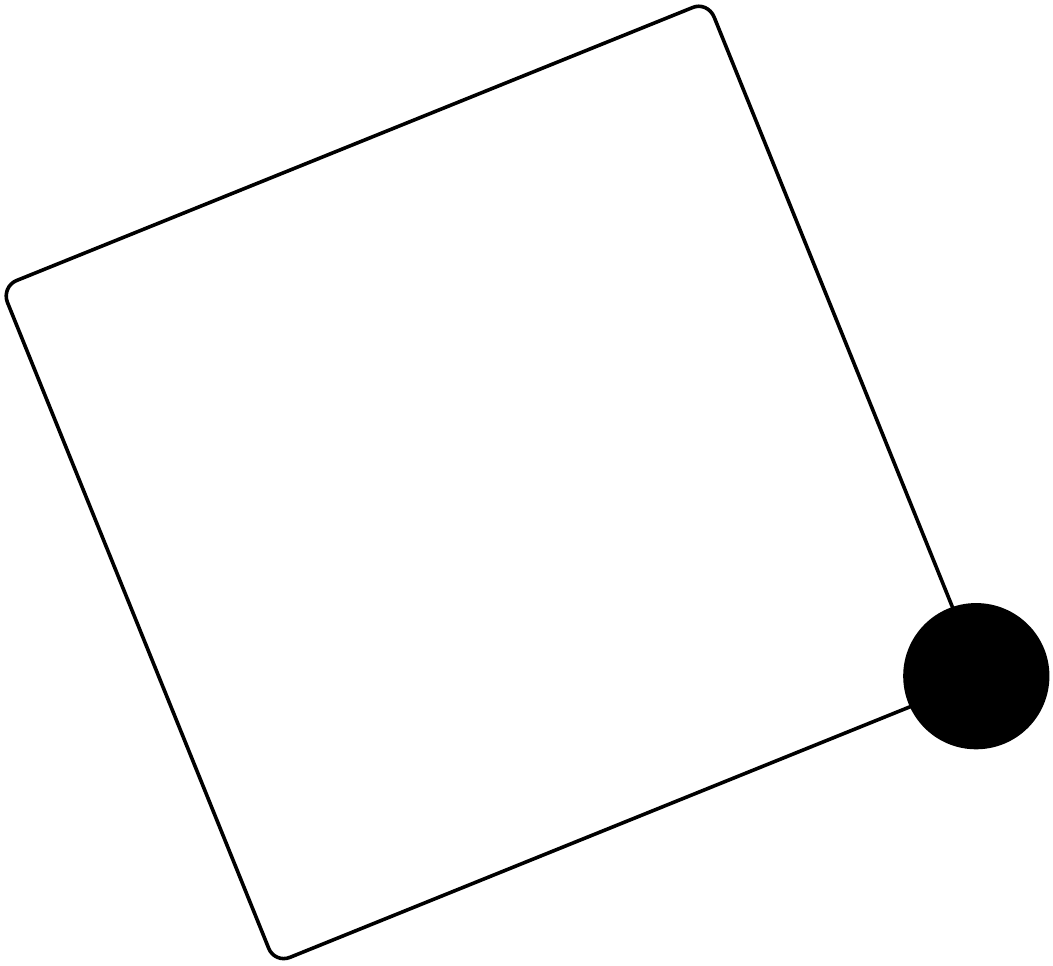} &   $+---$ & $\eps_1$ \\
			8 & \includegraphics[height=1cm,width=1cm]{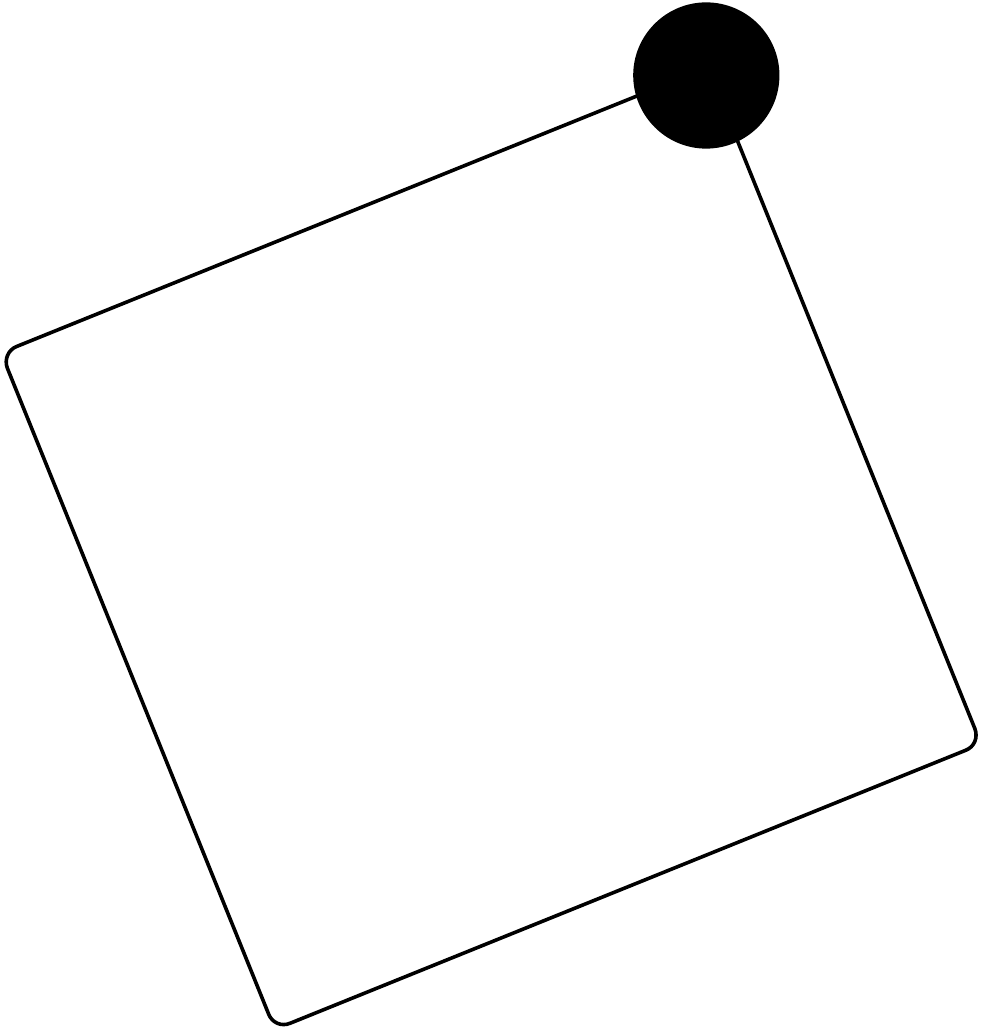} &  $-+--$ & \\
			9 & \includegraphics[height=1cm,width=1cm]{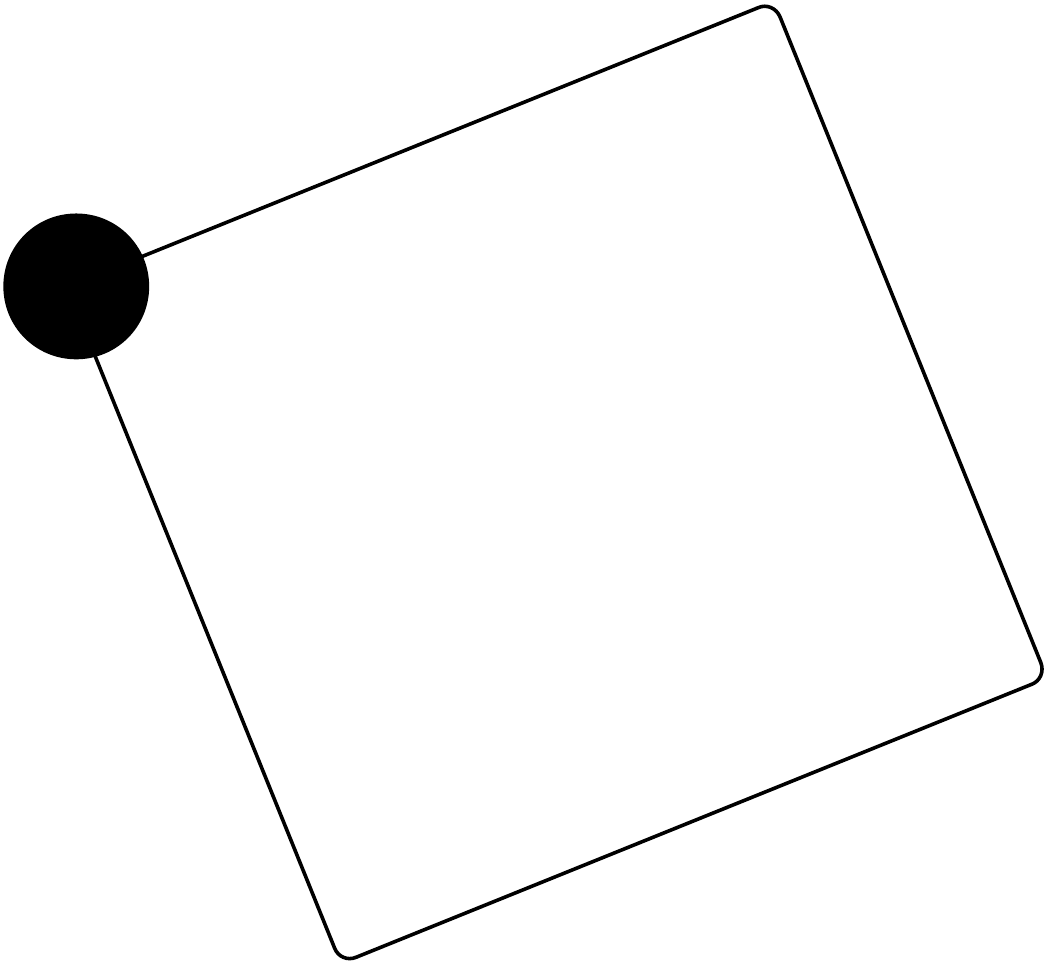} &  $--+-$ &  \\
			10&  \includegraphics[height=1cm,width=1cm]{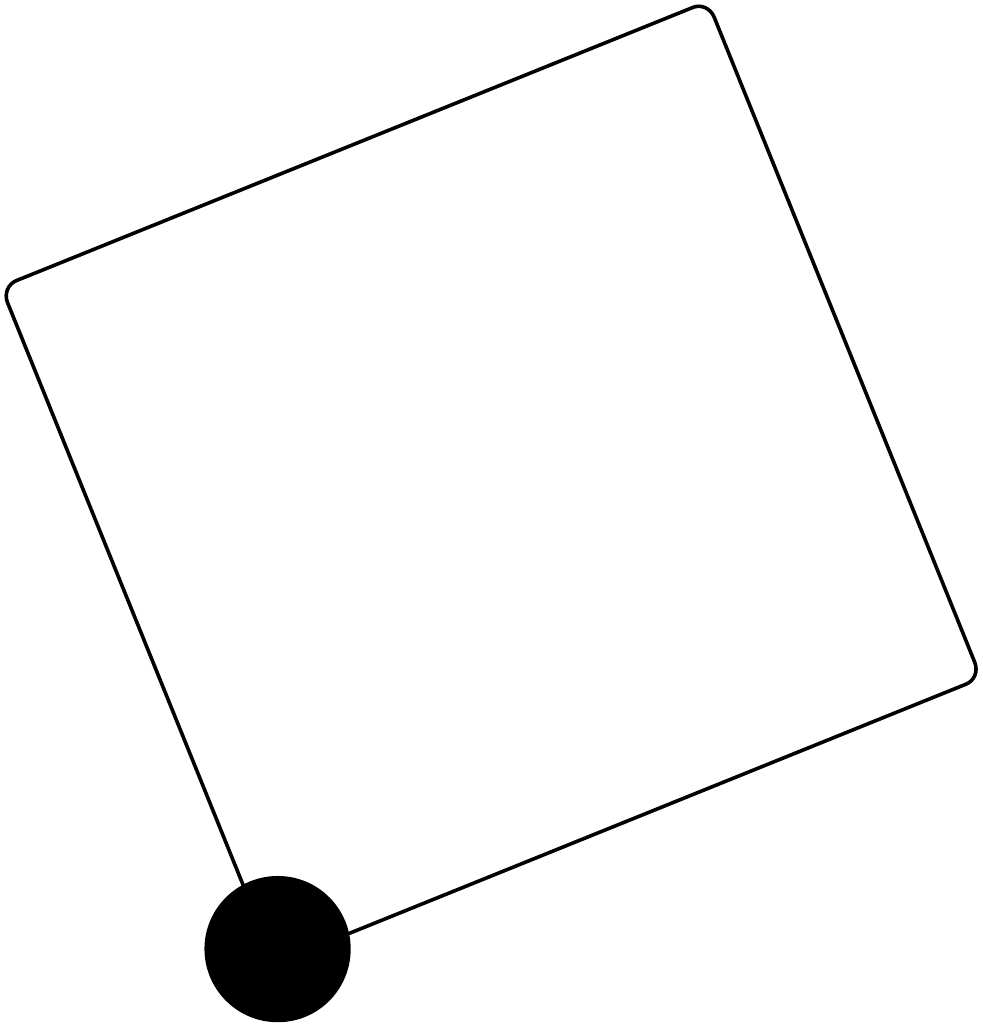} &   $---+$ & \\
			11 & \includegraphics[height=1cm,width=1cm]{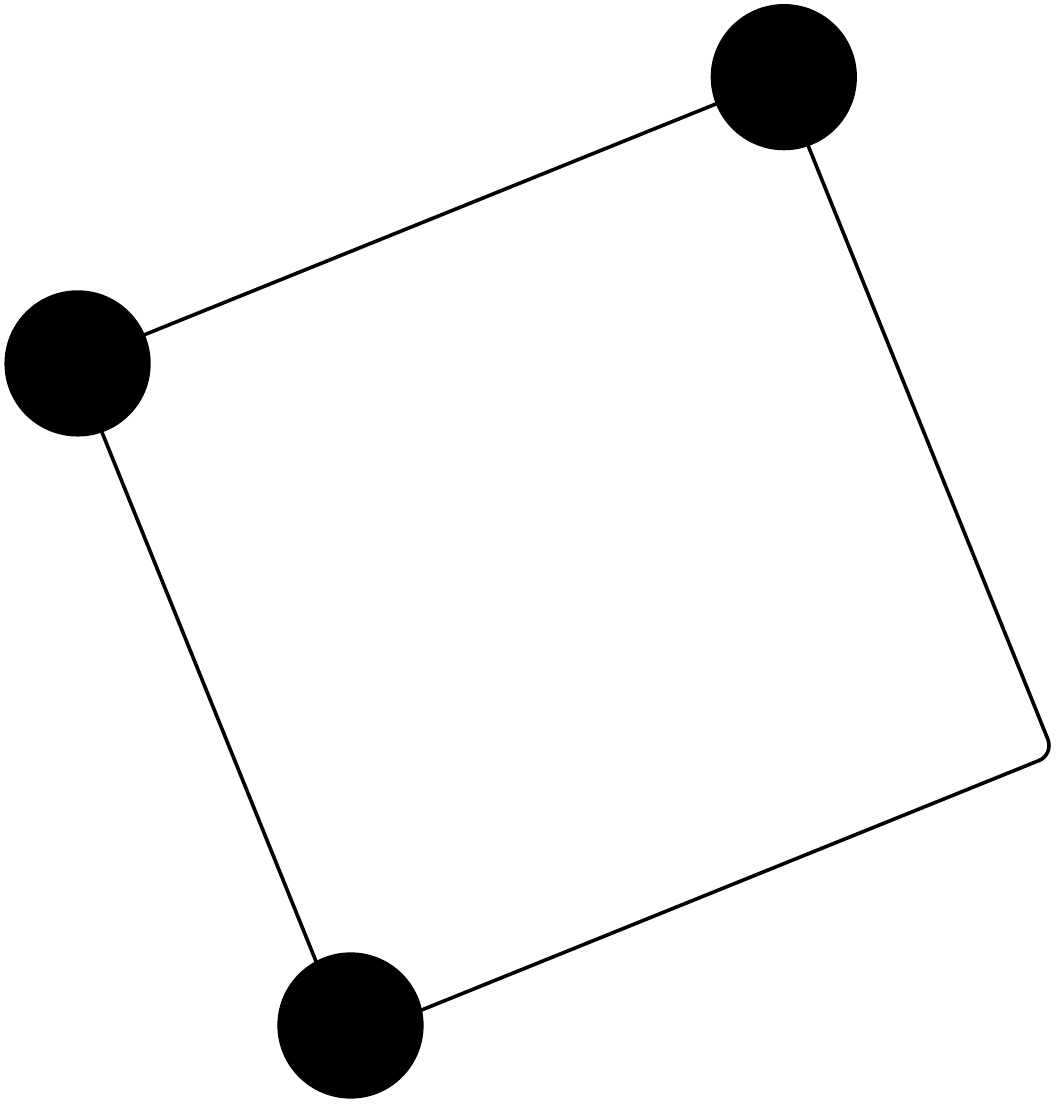} &  $-+++$ &  \\
			12 & \includegraphics[height=1cm,width=1cm]{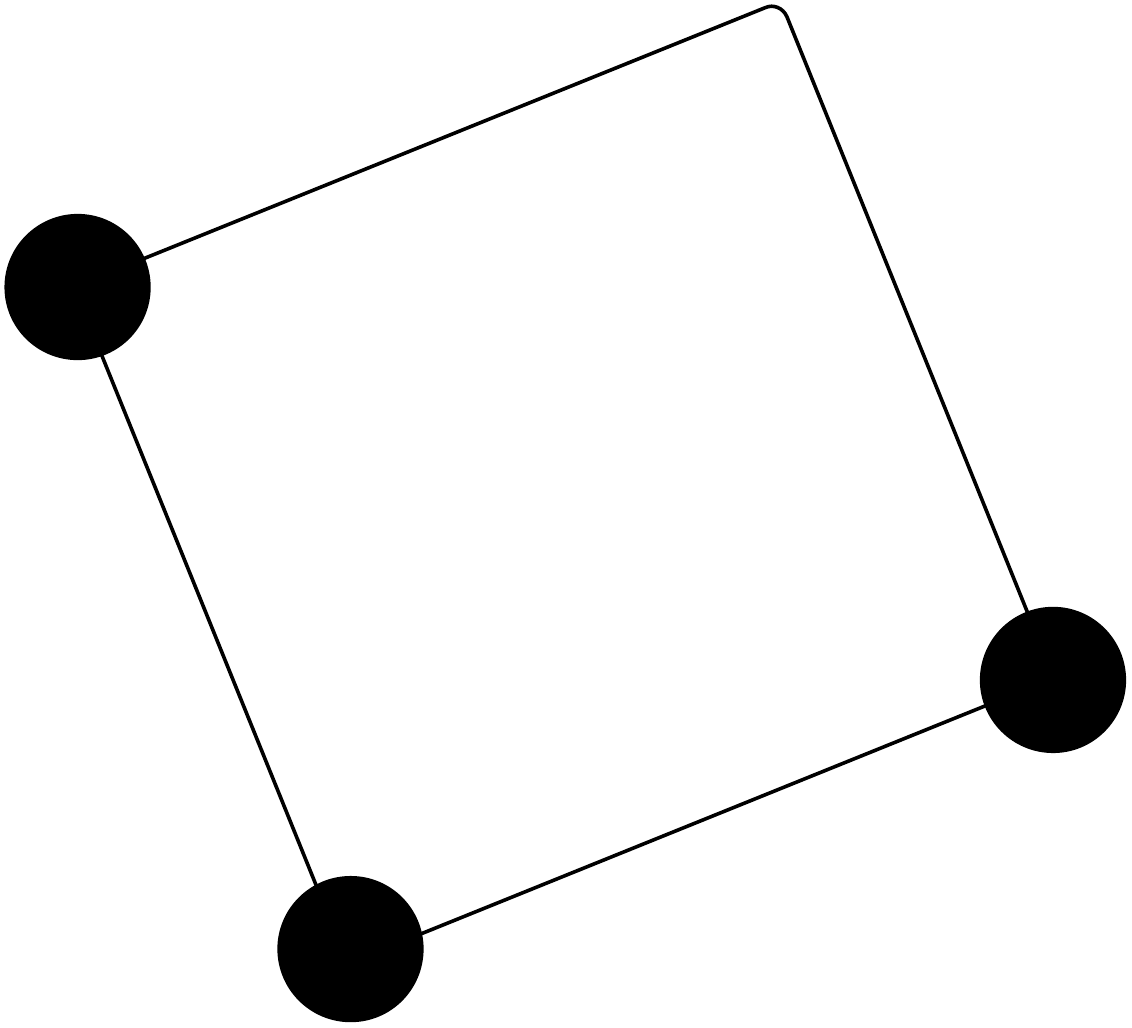} &   $+-++$&  \\
			13 & \includegraphics[height=1cm,width=1cm]{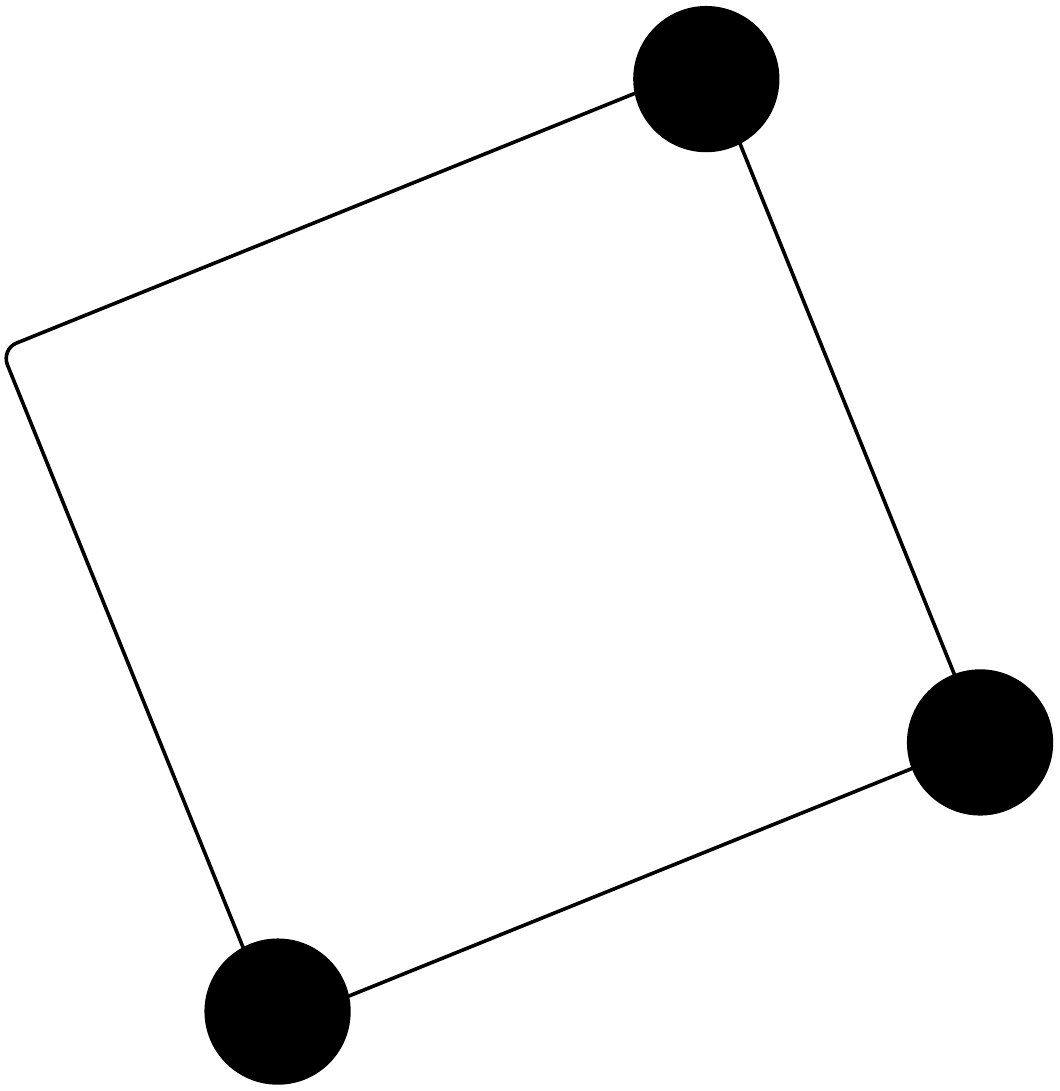}&   $++-+$& \\
			14 & \includegraphics[height=1cm,width=1cm]{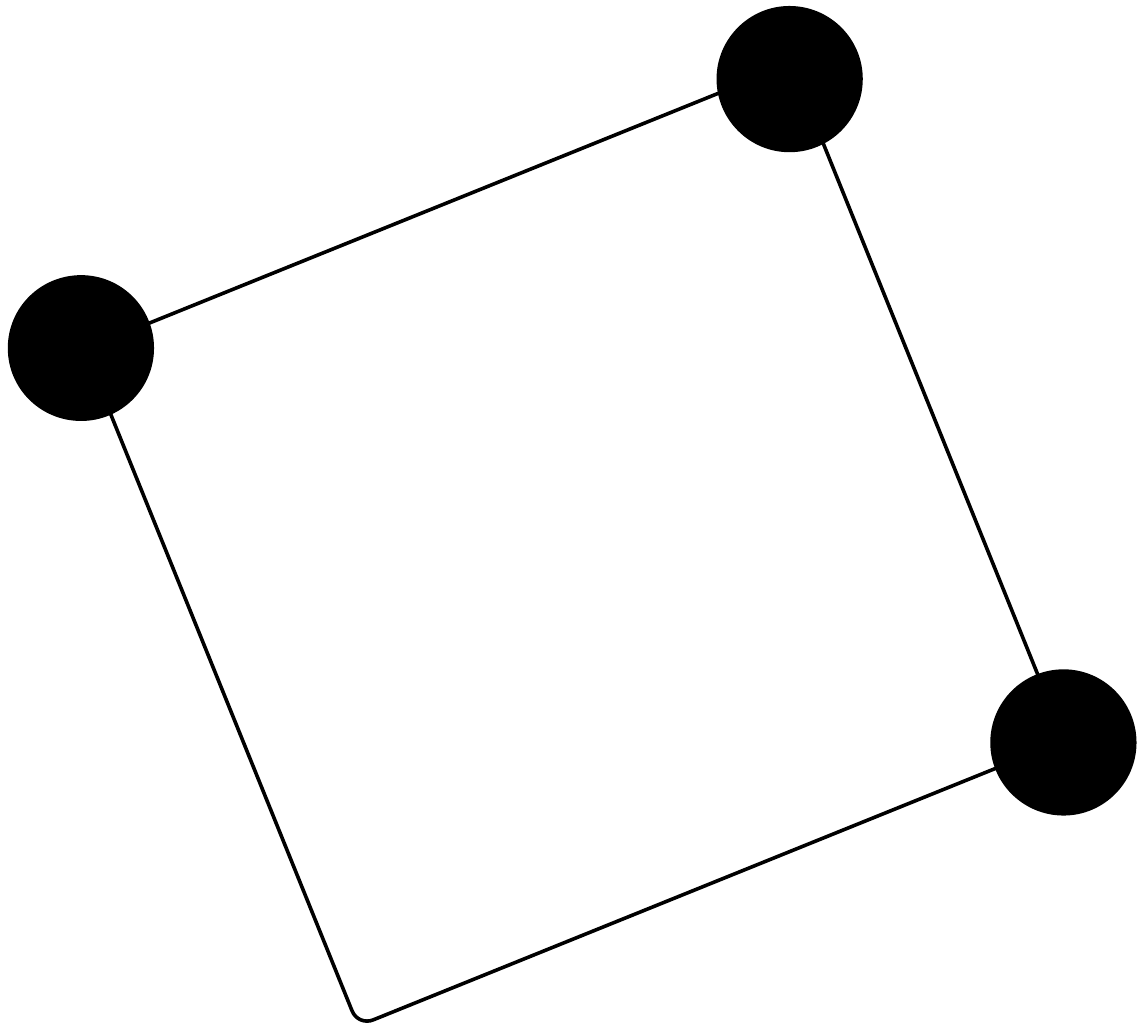} &   $+++-$ \\
			\hline
		\end{tabular}
	\end{center}
\end{table}

To proceed from the representation of proton configuration energies to the pseudospin representation, we use a standard procedure, originally developed for the KH$_2$PO$_4$ type crystals \cite{blinc:66,levitskii:04}, where the Hamiltonian of the
short-range correlations between protons, surrounding each A type C$_4$O$_4$ group 
is written as follows:
\be
\label{topseudospin}
H_{yq}^A = \sum_{i=1}^{16}\hat N_i(yq){\cal E}_{i}, \quad
\hat N_i(yq)=\prod\limits_{f=1}^4\frac{1}{2}(1+s_f{\sigma}_{yqf})
\ee
where $\hat N_i(yq)$ is the operator of the four-particle configuration $i$; $s_f=\pm1$ is the sign of the eigenvalue of the $\sigma_{yqf}$ operator in this particular configuration; ${\cal E}_{i}$ is the energy of the configuration. It is assumed
that $s_f=+1$ if a proton at the $f$th bond is localized at the H-site proximal to the particular A type C$_4$O$_4$ group, and $s_f=-1$ if the proton is localized at the other (distal) H-site of the same bond.  Using equation~(\ref{topseudospin}), we arrive at the following expression for the Hamiltonian 
\bea
&&
H_{yq}^A=V\left[\frac{\sigma_{yq1}}2\frac{\sigma_{yq2}}2+
\frac{\sigma_{yq2}}2\frac{\sigma_{yq3}}2+
\frac{\sigma_{yq3}}2\frac{\sigma_{yq4}}2+
\frac{\sigma_{yq4}}2\frac{\sigma_{yq1}}2\right]\nonumber\\
\label {H_short}&&  
\quad {} +   U\left[\frac{\sigma_{yq1}}2\frac{\sigma_{yq3}}2+
\frac{\sigma_{yq2}}2\frac{\sigma_{yq4}}2\right]
+\Phi \frac{\sigma_{yq1}}2\frac{\sigma_{yq2}}2\frac{\sigma_{yq3}}2\frac{\sigma_{yq4}}2.
\eea
The short-range interaction constants
\be
\label{Slater-ham} V=-\frac{\eps-w_{1}}{2},\quad
U=\frac{\eps+w_{1}}{2},\quad \Phi=2\varepsilon-8w+2w_{1},
\ee
are linear functions of the Slater-Takagi type energy parameters
\[
\eps=\eps_s-\eps_a, \quad w=\eps_1-\eps_a, \quad w_1=\eps_0-\eps_a.
\]
Note that the model of non-interacting perpendicular one-dimensional chains is obtained from equation~(\ref{H4}) at $\Phi=V=0$, i.e. at $\eps=w_1=2w$, where the following order of the configuration energies should be assumed \cite{maier:82} $\eps_a<\eps_1<\eps_s=\eps_0$.

It can be shown, as it has been done for the KH$_2$PO$_4$ ferroelectrics, that the contributions of the correlations from the A and B type groups
to the total thermodynamic potential are equal. The Hamiltonian of the short-range interactions in this case can be written as follows:
\begin{equation}H_{\textrm{short}} \rightarrow 2\sum_{qy}H^A_{qy}, \end{equation}
where the expression for $H^A_{qy}$ is given by equation~(\ref{H_short}).

The long-range interactions in the system Hamiltonian (\ref{sqa-Ham}) are the dipole-dipole interactions, in analogy to the case of KH$_2$PO$_4$ \cite{blinc:66},  renormalized by the proton-lattice coupling. The MFA is successfully used for these interactions in the calculations for  hydrogen-bonded ferroelectrics, if the  fluctuations in the vicinity of the phase transition temperature neglected in this approximation are not the subject of special interest.

Within the MFA, we obtain the following expressions for the long-range intralayer 
\begin{equation}\label{intra}
H_{\textrm{long}}^{\textrm{intra}}=-\frac12\sum_{y=1}^{N_y}\sum_{qq' \atop ff'}
J^{\textrm{intra}}_{ff'}(qq')\frac{\sigma_{yqf}}{2}\frac{\sigma_{yq'f'}}{2}\simeq
-2\sum_{yqf}F_{yqf}^{\textrm{intra}}\frac{\sigma_{yqf}}{2}+\sum_{yqf}F_{yqf}^{\textrm{intra}}\frac{\langle\sigma_{yqf}\rangle}{2}
\end{equation}
and for interlayer 
\begin{equation}
H_{\textrm{long}}^{\textrm{inter}}=-\frac12\sum_{y}\sum_{y'\neq y}\sum_{qq' \atop ff'}
J^{\textrm{inter}}_{ff'}(yy';qq')\frac{\sigma_{yqf}}{2}\frac{\sigma_{y'q'f'}}{2}\simeq
-2\sum_{yqf}F_{yqf}^{\textrm{inter}}\frac{\sigma_{yqf}}{2}+\sum_{yqf}F_{yqf}^{\textrm{inter}}\frac{\langle\sigma_{yqf}\rangle}{2}
\end{equation}
interactions. Here $N_y$ is the total number of the layers.
%Here, $y$ stands for the layer index, $N_y$ is the total number of the layers,
%$q$ is the index of the A type C$_4$O$_4$ group, and $f$ is the bond index. 
The internal mean fields are as follows:
%\begin{equation}
%\label{fields1}
%F_{yqf}^{\textrm{intra}}=\frac 14\sum_{q'f'}J^{\textrm{intra}}_{ff'}(qq'),\quad
%F_{yqf}^{\textrm{inter}}=\frac 14\sum_{y'q'f'}J^{\textrm{intra}}_{ff'}(yy;qq').
%\end{equation}
\begin{equation}
F_{yqf}^{\textrm{intra}}=\frac
14\sum_{q'f'}J^{\textrm{intra}}_{ff'}(qq')\langle \sigma_{yq'f'}
\rangle,\quad F_{yqf}^{\textrm{inter}}=\frac
14\sum_{y'q'f'}J^{\textrm{intra}}_{ff'}(yy;qq')\langle \sigma_{y'q'f'}
\rangle.
\label{fields1}
\end{equation}

The following symmetry of the pseudospin mean values is assumed for the antiferroelectrically ordered two-sublattice model in the absence of external electric field
\begin{equation}\label{symmetry1}
\langle \sigma_{yqf} \rangle=\exp[\ri{\bf k}_2 {\bf R}_y] \eta_{f}.
\end{equation}
Here, ${\bf k}_2=(0,b_2,0)$; $b_2$ is the basic vector of
the reciprocal lattice; the factor $\exp[\ri{\bf k}_2 {\bf R}_y]=\pm 1$ denotes two sublattices of an antiferroelectric,
 ${\bf R}_y$ is the position vector of the $y$-th layer, and 
\begin{equation}\label{symmetry2}
 \eta_{1}=-\eta_3, \quad \eta_{2}=-\eta_4, \quad \eta_{1}\approx\eta_2.
\end{equation}
The last relation reflects the slight non-equivalence, mentioned in introduction,  of hydrogen bonds, linking C$_4$O$_4$ groups along the $a$ and $c$ axes.

Even though each particular interaction parameter  $J^{\textrm{intra}}_{ff'}(qq')$ and $J^{\textrm{inter}}_{ff'}(yy';qq')$ is obviously changed by the strains $\eps_1$ and $\eps_3$,
the symmetry of the long-range interaction matrices \textit{Fourier transforms}
\[
J^{\textrm{intra}}_{ff'}(0)=\sum_{q'}J^{\textrm{intra}}_{ff'}(qq'),\quad
J^{\textrm{inter}}_{ff'}({\bf k}_2)=\sum_{q-q'}\sum_{y-y'}J^{\textrm{inter}}_{ff'}(yy';qq')\exp[\ri{\bf k}_2 ({\bf R}_y-{\bf R}_{y'})]
\]
over the bond indices $f$ and $f'$, as can be easily checked, remains unchanged in the presence of the orthorhombic strain $\eps_1-\eps_3$:
\begin{eqnarray}
\label{symmetry3}
 J_{11}=J_{22}=J_{33}=J_{44},\quad 
\label{Jsymm} J_{12}=J_{23}=J_{34}=J_{41},\quad J_{13}=J_{24}
\end{eqnarray}
both for $J^{\textrm{intra}}(0)$ and $J^{\textrm{inter}}_{ff'}({\bf k}_2)$. 
The symmetry (\ref{symmetry3}) is obvious for strictly square-shaped C$_4$O$_4$ groups (point group C$_{4h}$). This is a statistically average symmetry of the paraelectric phase, where the hydrogens are placed, also statistically, in the middle of the hydrogen bonds.

Taking into account equations~(\ref{symmetry1})--(\ref{symmetry3}), we can write the Hamiltonians of the long-range interactions as follows:
\begin{equation}
\label{Hlong}
H_{\textrm{long}}=H^{\textrm{intra}}_{\textrm{long}}+
H^{\textrm{inter}}_{\textrm{long}}
=N\nu[\eta_1^2+\eta_2^2]-2\nu\sum_{yq}\exp[\ri{\bf k}_2 {\bf R}_y]\left[
\eta_1\frac{\sigma_{qy1}-\sigma_{qy3}}{2}+
\eta_2\frac{\sigma_{qy2}-\sigma_{qy4}}{2}\right],
\end{equation}
where
\begin{equation}
\nu=\nu^{\textrm{intra}}(0)+
\nu^{\textrm{inter}}({\bf k}_2)
=\frac{J_{11}^{\textrm{intra}}(0)-J_{13}^{\textrm{intra}}(0)}4+
\frac{J_{11}^{\textrm{inter}}({\bf k}_2)-J_{13}^{\textrm{inter}}({\bf k}_2)}4.
\end{equation}
We also took into account the fact that $N_yN_{qA}=N$.

For the sake of simplicity,  we shall hereafter neglect the weak non-equivalence of the perpendicular chains of hydrogen bonds and use a single order parameter
\begin{equation}\label{symmetry4}
\eta\equiv\eta_{1}=\eta_2=-\eta_3=-\eta_4
\end{equation}
instead of equation~~(\ref{symmetry2}).

The four-particle cluster approximation will be used for the short-range interactions, described by the Hamiltonian (\ref{H_short}). With the long-range interactions taken into account in the mean field approximation, the thermodynamic potential of the system should be written as follows:
\begin{eqnarray}
 G&=& -vN\sum_{i=1}^3\sigma_i\eps_i+ NU_{\textrm{seed}} -\frac1\beta\sum_{qy}\left[2\ln \textrm{Sp} \exp (-\beta H_{qy}^{(4)}) -\sum_{f=1}^4 \ln \textrm{Sp} \exp (-\beta H_{qyf}^{(1)})\right] \nonumber\\
&+&\sum_{yqf}\left(F_{yqf}^{\textrm{intra}}+F_{yqf}^{\textrm{inter}}\right)
\frac{\langle\sigma_{yqf}\rangle}{2}
\label{tpot}.
\end{eqnarray}
Here $\sigma_1=\sigma_2=\sigma_3=-p$, $\,\,\beta=(k_{\textrm{B}}T)^{-1}$, and
\begin{equation}
F_{yq1}^{\textrm{intra}}+F_{yq1}^{\textrm{inter}}=
F_{yq2}^{\textrm{intra}}+F_{yq2}^{\textrm{inter}}=
-F_{yq3}^{\textrm{intra}}-F_{yq3}^{\textrm{inter}}=
-F_{yq4}^{\textrm{intra}}-F_{yq4}^{\textrm{inter}}=
\exp[\ri{\bf k}_2 {\bf R}_y]\nu\eta.
\end{equation}
The four-particle cluster Hamiltonian is
\be
\label{H4}
H_{qy}^{(4)}=H^A_{qy}-\sum_{f=1}^4\frac{z_{yqf}}{\beta}\frac{\sigma_{yqf}}2\,,
\ee
where
\[
z_{yqf}=\beta(\Delta_{yqf}+2F_{yqf}^{\textrm{intra}}+2F_{yqf}^{\textrm{inter}}).
\]
%$\beta=(k_{\textrm{B}}T)^{-1}$. 
The fields $\Delta_{yqf}$ are the effective cluster fields that describe short-range interactions of the spin $\sigma_{yqf}$ with the particles from outside the cluster $q$. They are determined from the self-consistency condition stating that pseudospin mean values calculated with the four-particle  (\ref{H4}) and with the one-particle
\[ H^{(1)}_{yqf}=-\left[{2\Delta_{yqf}+2F_{yqf}^{\textrm{intra}}+2F_{yqf}^{\textrm{inter}}}{}\right]\frac{\sigma_{yqf}}2\] 
Hamiltonians should coincide. We get
\begin{equation}
\label{z}
z_{yq1}=z_{yq2}=-z_{yq3}=-z_{yq4}=
\exp[\ri{\bf k}_2 {\bf R}_y]z\,, \quad z=\frac12\ln\frac{1+\eta}{1-\eta}+\beta\nu\eta.
\end{equation}

Taking into account equations (\ref{Hlong}), 
(\ref{symmetry4}), (\ref{H4}), (\ref{z}), the thermodynamic potential per one unit cell is obtained in the following form
\begin{equation}\label{sqa:tpot}
g=U_{\textrm{seed}} -\frac2\beta\left[ \ln D +\ln(1-\eta^2)\right]  +2\nu\eta^2 -v\sum_{i=1}^3\sigma_i\eps_i\,,
\end{equation}
where
\[D=a+\cosh2z+4b\cosh z+1,\quad 
a=\exp(-\beta\eps), \quad b=\exp(-\beta w).
\]

In the earlier theories \cite{stasyuk:99,torstveit}, the short-range Slater-Takagi energies in the KH$_2$PO$_4$ family crystals were considered as quadratic functions of the distance $\delta$. For the squaric acid, we shall employ the same scheme. Using the term of the relative deviation of $\delta$ from its value $\delta_0$ at ambient pressure (we shall call it a displacement $\mu'$)
\begin{equation}
\mu'=\frac{\delta-\delta_0}{\delta_0}\,,
\end{equation}  
we assume that
\begin{equation}
\label{kdp-Slater}
\eps=\eps_0(1+\mu')^2\,, \quad w=w_0(1+\mu')^2\,.
\end{equation}
Here, the quadratic in $\mu'$ terms, omitted in \cite{stasyuk:99},  are included into consideration.

For the parameter of long-range (dipole-dipole) interactions $\nu$, we take into account both the dependence of the dipole moments on $\delta$ and the changes in the interaction parameter due to the overall crystal deformation \cite{stasyuk:99} and associated with changes in the equilibrium distances between protons (dipoles) 
\begin{equation}
\label{kdp-long}
\nu=\nu_0(1+\mu')^2+\sum_{i=1}^3\psi_i\eps_i.
\end{equation}
It should be underlined that none of the earlier theories \cite{stasyuk:99,torstveit} described the thermal expansion of the crystals; therefore, the deformational effects there were only those caused by external pressures. On the contrary, in the present model the strains $\eps_i$ are induced both by temperature changes and by external pressures. In the mean field approximation, the expansion (\ref{kdp-long}) gives rise to the terms of  electrostriction type in the
Hamiltonian, linear in the strains and quadratic in the sublattice polarization (order parameter~$\eta$).

In \cite{stasyuk:99,torstveit}, the distance $\delta$ was treated as a pressure dependent and temperature independent model parameter, with the linear pressure dependence chosen either from the available experimental data or by fitting the theory to experiment for the transition temperatures. In the present work, we shall use a similar approach and assume $\delta$ to vary according to its experimentally observed \textit{above the transition} linear temperature \cite{semmingsen:95} and external hydrostatic pressure $p$ \cite{mcmahon:90} dependences
\begin{equation}
\label{delta-model}
\delta=\delta_0[1+\delta_pp+\delta_T(T-T_{\textrm{N0}})],
\end{equation}
where $T_{\textrm N0}$ is the transition temperature at ambient pressure.
It means that the anomalous temperature behavior of $\delta$ below the transition point and its jump at $T_{\textrm N}$ are neglected. 

Minimization of the thermodynamic potential (\ref{sqa:tpot}) with respect to the order parameter $\eta$ and strains~$\eps_i$
\[
\frac{\partial g}{\partial \eta}=0,\quad
\frac{\partial g}{\partial \eps_i}=0
\]
 yields the following equations
\begin{eqnarray*}
&&\eta=\frac{\sinh 2z+2b\sinh z}{D},\nonumber\\
&&\sigma_i=\sum_{j=1}^3c_{ij}^{(0)}\eps_j-\sum_{j=1}^3c_{ij}^{(0)}\alpha_j^{(0)}(T-T_j^0)+\frac{2\psi_i\eta}v\left(\eta-2\frac{\sinh 2z+2b\sinh z}{D}\right).
\end{eqnarray*}
From the above it follows that in equilibrium
\begin{equation}
\eps_k=\alpha_k^{(0)}(T-T_k^0)+\sum_{i=1}^3 \sigma_i s_{ki}^{(0)} -\frac{2\eta^2}v
	\sum_{i=1}^3 \psi_i s_{ki}^{(0)},
\label{eps-anom}	
\end{equation}
where $s_{ki}^{(0)}$ is the matrix of ``seed'' elastic compliances, inverse to $c_{ij}^{(0)}$. One can see that in the paraelectric phase ($\eta=0$), the microscopic contributions to the strains vanish, whereas in the ordered phase they are proportional to $\eta^2$, indicating the electrostriction type contributions  governed by the parameters $\psi_i$.

Finally, the molar entropy of the proton subsystem is as follows:
\begin{equation}
\label{entropy}
\Delta S=\ln[(1-\eta^2)D] -\frac{1}{DT}\left[a\varepsilon+4bw\cosh z\right]-2\frac{\nu\eta^2}{T}.
\end{equation}

\section{Calculations}
\label{calculations}
In the calculations, the thermodynamic potential is minimized numerically
with respect to the order parameter $\eta$. At the same time, the strains $\eps_i$ are determined from equation~(\ref{eps-anom}). 

The quantities that should be described include: \\
the temperature curves at ambient pressure of
\begin{itemize}
	\item the sublattice polarization (order) parameter,
	\item macroscopic lattice strains $\eps_i$,
	\item thermal expansion coefficients and specific heat;
\end{itemize}
the pressure curves of
\begin{itemize}
	\item the transition temperature $T_{\textrm N}$,
	\item lattice strains $\eps_i$.
\end{itemize}

Since we do not want to overcomplicate the fitting procedure by adopting different values of the model parameters for  paraelectric and antiferroelectric phase, the chosen matrix quantities $c_{ij}^{(0)}$, 
$\alpha_i^{(0)}$, $T_i^0$ should obey the tetragonal symmetry of  paraelectric phase, namely
$c_{11}^{(0)}=c_{33}^{(0)}$, $c_{12}^{(0)}=c_{23}^{(0)}$, $\alpha_1^{(0)}=\alpha_3^{(0)}$, $T_1^0=T_3^0$.

There are four different ``seed'' elastic constants  $c_{11}^{(0)}$, $c_{22}^{(0)}$,
$c_{12}^{(0)}$, and $c_{13}^{(0)}$. We take $c_{11}^{(0)}$ to be equal to
the experimental value of $c_{11}$ \cite{rehwald:78} above the transition point.
Experimental elastic constant $c_{22}$ was found to slightly decrease with an increasing temperature \cite{rehwald:78,yamanaka:87}. The ``seed'' $c_{22}^{(0)}$ is chosen accordingly, coinciding with the data of \cite{rehwald:78}. Finally, $c_{12}^{(0)}$ and $c_{13}^{(0)}$, for which no convincing experimental data are available, were chosen to provide a correct fit to the experimental  \cite{katrusiak:86} pressure dependence of the lattice constants $a$ and $b$ at 292~K.

The parameters of the short-range correlations $\eps_0$ and $w_0$ govern the temperature behavior of the order parameter $\eta$ (in particular, the magnitude of its jump at the transition $\Delta \eta_c$ and steepness of its rise to saturation with lowering temperature) and the value of the transition temperature at ambient pressure $T_{\textrm N0}$. The latter is also extremely sensitive to the value of the long-range interactions parameter $\nu_0$. Hence, the set of $\eps_0$, $w_0$, and $\nu_0$ is chosen to yield $T_{\textrm N0}=373.5$~K, $\Delta \eta_c\approx 0.57$, as well as a correct temperature curve of $\eta$ between the transition and saturation. Contributions of the double-ionized configurations are neglected by putting $w_1\to\infty$. 

The ``seed'' thermal expansion coefficients $\alpha_i^{(0)}$  as well as the parameters $\psi_i$ are determined by fitting the theoretical temperature dependences of the lattice strains  to experimental data \cite{ehses:81,johansen:84}. In fact, $\alpha_i^{(0)}$ should be simply equal to the corresponding paraelectric experimental values, as is seen directly in equation~(\ref{eps-anom}). The parameters $\psi_i$, on the other hand, are unambiguously determined by fitting the calculated anomalous parts of the strains to the experiment below the transition temperature, using the same equation~(\ref{eps-anom}). 
As $\psi_i$ are relevant for the ordered phase only, they do not have to adhere to the symmetry of the paraelectric phase; hence, we can take $\psi_1\neq\psi_3$, as is indeed required by the just described fitting.

As we have already mentioned, the temperatures $T_i^0$ determine the reference point of the thermal expansion of the crystal, which can be set arbitrarily. Thus, $T_i^0$ are not fitting parameters of the model and, therefore, can also  be chosen arbitrarily. In our calculations we chose them to yield zero values of the lattice strains $\eps_i$ just above the transition temperature at ambient pressure. In fact, $T_1^0=T_2^0=T_{\textrm N0}=373.5$~K, as seen from equation~(\ref{eps-anom}).

As already described, we take $\delta$ to vary according to its experimentally observed linear temperature  and external hydrostatic pressure (\ref{delta-model}). The coefficients $\delta_T$ and $\delta_p$ are deduced from the data of \cite{semmingsen:95} and~\cite{mcmahon:90}.

The unit cell volume is $v=2\cdot10^{-28}$~m$^3$ just above the transition
point at ambient pressure, as determined from the data of \cite{ehses:81}.
The final values of the model parameters, used in our calculations, are
given in table~\ref{tbl1}.

%\begin{table}[!b]
%	\caption{The adopted values of the model parameters. }
%	\label{tbl1}
%	\vspace{2ex}
%	\begin{center}
%		\renewcommand{\arraystretch}{0}
%%
%\begin{tabular}{cccccc|cc|cc}
%	\hline
%	 $\eps_0/k_{\text{B}}$ & $w_0/k_{\text{B}}$ &  $\nu_0/k_{\text{B}}$ & $\psi_1/k_{\text{B}}$ & $\psi_2/k_{\text{B}}$ & $\psi_3/k_{\text{B}}$ & $\alpha_1^0$ & $\alpha_2^0$ &$\delta_T$ & $\delta_p$ \\
%	   \multicolumn{6}{c|}{K} & \multicolumn{2}{c|}{$10^{-5}$ K$^{-1}$} & $10^{-4}$ K$^{-1}$ & kbar$^{-1}$ \\
%	\cline{1-10}
% 395 &  1100 & 79.8 & $-518$
%	& 445 & 1096 & 1.2 & 13.0 & 2 & $-0.014$ \strut 	\\ 	\hline
%\end{tabular}
%	\vspace{1ex}
%\begin{tabular}{cccc}
%	\hline
%% & \\this work
% %\\ \hline
%	 $c_{11}^{0}$&  $c_{12}^{0}$ & $c_{13}^{0}$ & $c_{22}^{0}$  \\
%	 \multicolumn{4}{c}{$10^{10}$ N/m$^{2}$}\\
%	\cline{1-4}\\
%	 6.5 & 2.3 & $-3.1$ & $2.38{-}0.02T$  \strut
%	\\
%	\hline
%\end{tabular}
%\renewcommand{\arraystretch}{1}
%	\end{center}
%\end{table}
%
\begin{table}[!b]
	\caption{The adopted values of the model parameters. }
	\small
	\label{tbl1}
	\vspace{2ex}
%	\begin{center}
		\renewcommand{\arraystretch}{0}
\begin{tabular}{cccccc|cc|cc| cccc}
	\hline
	 $\eps_0/k_{\text{B}}$ & $w_0/k_{\text{B}}$ &  $\nu_0/k_{\text{B}}$ & $\psi_1/k_{\text{B}}$ & $\psi_2/k_{\text{B}}$ & $\psi_3/k_{\text{B}}$ & $\alpha_1^0$ & $\alpha_2^0$ &$\delta_T$ & $\delta_p$    &$c_{11}^{0}$&  $c_{12}^{0}$ & $c_{13}^{0}$ & $c_{22}^{0}$  \\
	   \multicolumn{6}{c|}{K} & \multicolumn{2}{c|}{$10^{-5}$ K$^{-1}$} & $10^{-4}$ K$^{-1}$ & kbar$^{-1}$ & \multicolumn{4}{c}  
	    {$10^{10}$ N/m$^{2}$}\strut\\ 
	\cline{1-14}
\strut
 395 &  1100 & 79.8 & $-518$
	& 445 & 1096 & 1.2 & 13.0 & 2 & $-0.014$ & 6.5 & 2.3 & $-3.1$ & $2.38{-}0.02T$\strut  \\	\hline
%	\end{center}
\end{tabular}
\end{table}

\begin{figure}[!t]
	\centerline{\includegraphics[height=0.3\textwidth]{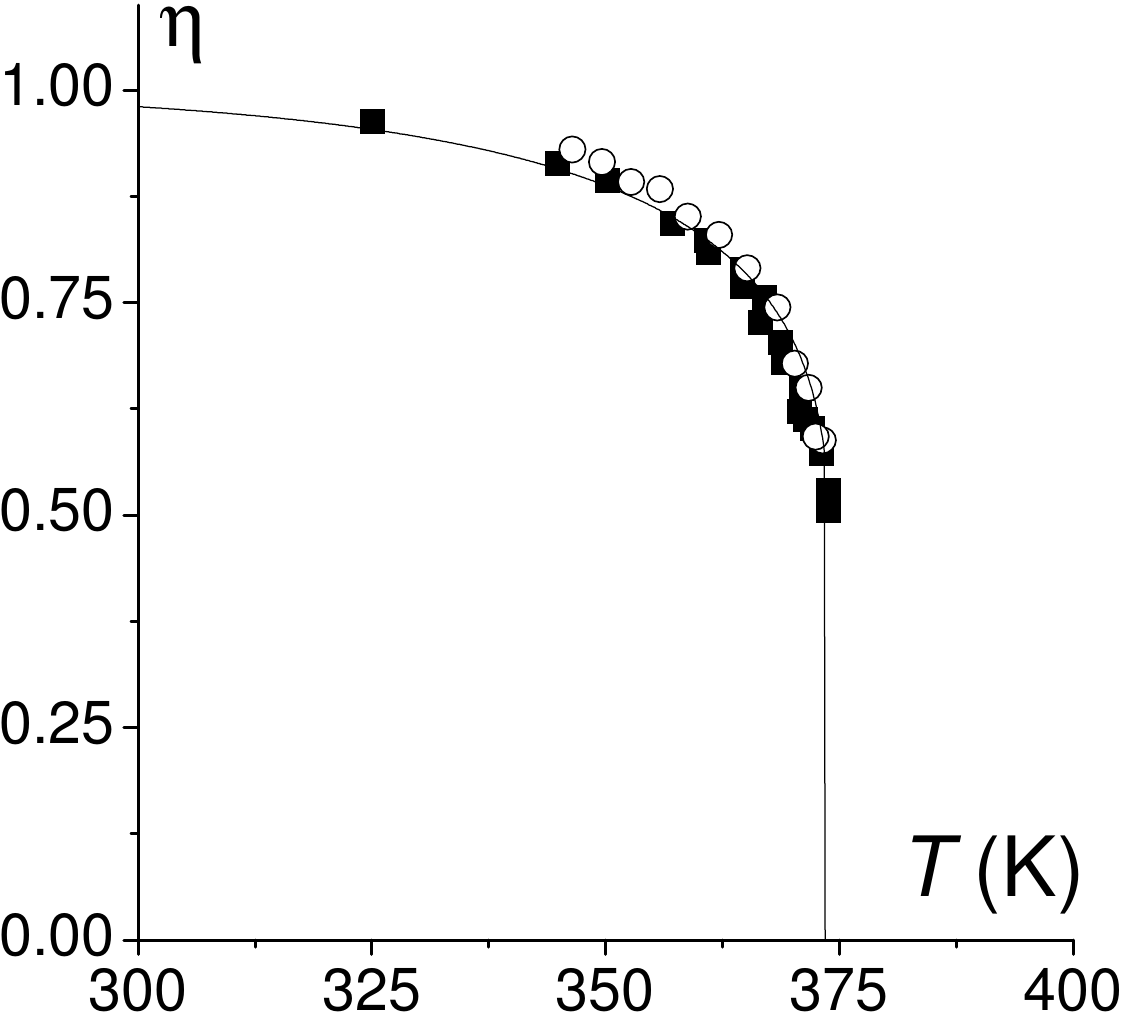}\hspace{5ex}
	\includegraphics[height=0.3\textwidth]{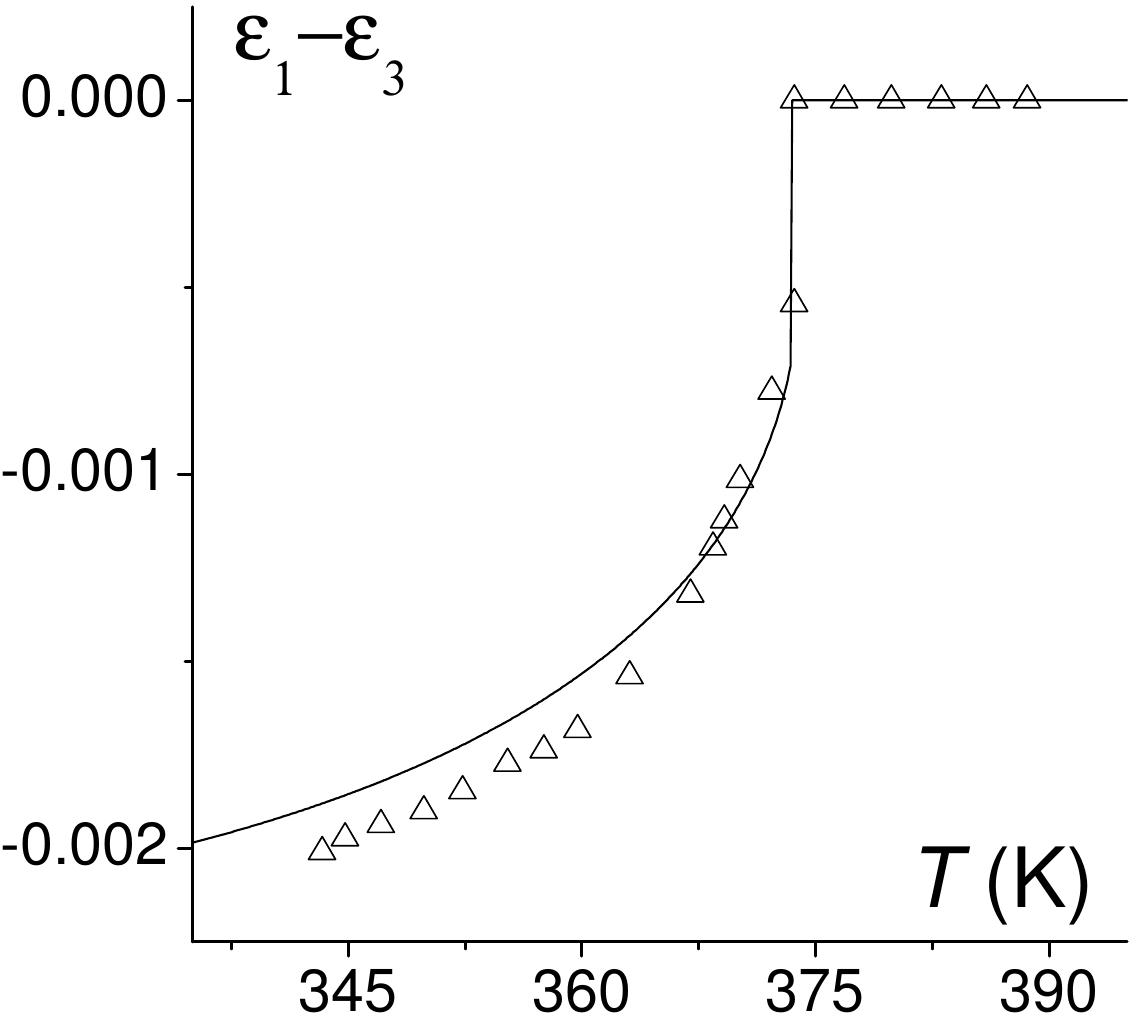}}
	\caption{Temperature dependence of the order parameter and spontaneous strain $\eps_1-\eps_3$ of squaric acid. Lines: the theory; symbols are experimental points taken from  \cite{klymachyov:97} ($\blacksquare$),   \cite{mehring:81} ($\bigcirc$), and \cite{ehses:81} ($\bigtriangleup$). } \label{fig-ordpar}
\end{figure}

In figure~\ref{fig-ordpar} we show the calculated temperature dependence of the order parameter $\eta$ and the spontaneous strain $\eps_1-\eps_3$ at ambient pressure. Experimental points for $\eta$ were obtained from the ${}^{13}$C NMR measurements. A clear first order phase transition is observed, with the jump of the order parameter $\Delta\eta_c\simeq0.57$. The spontaneous strain $\eps_1-\eps_3$ is negative below the transition and, as it follows from equation~(\ref{eps-anom}), it  is proportional to the square of the order parameter $\eta^2$.

\begin{figure}[htb]
	\centerline{\includegraphics[height=0.45\textwidth]{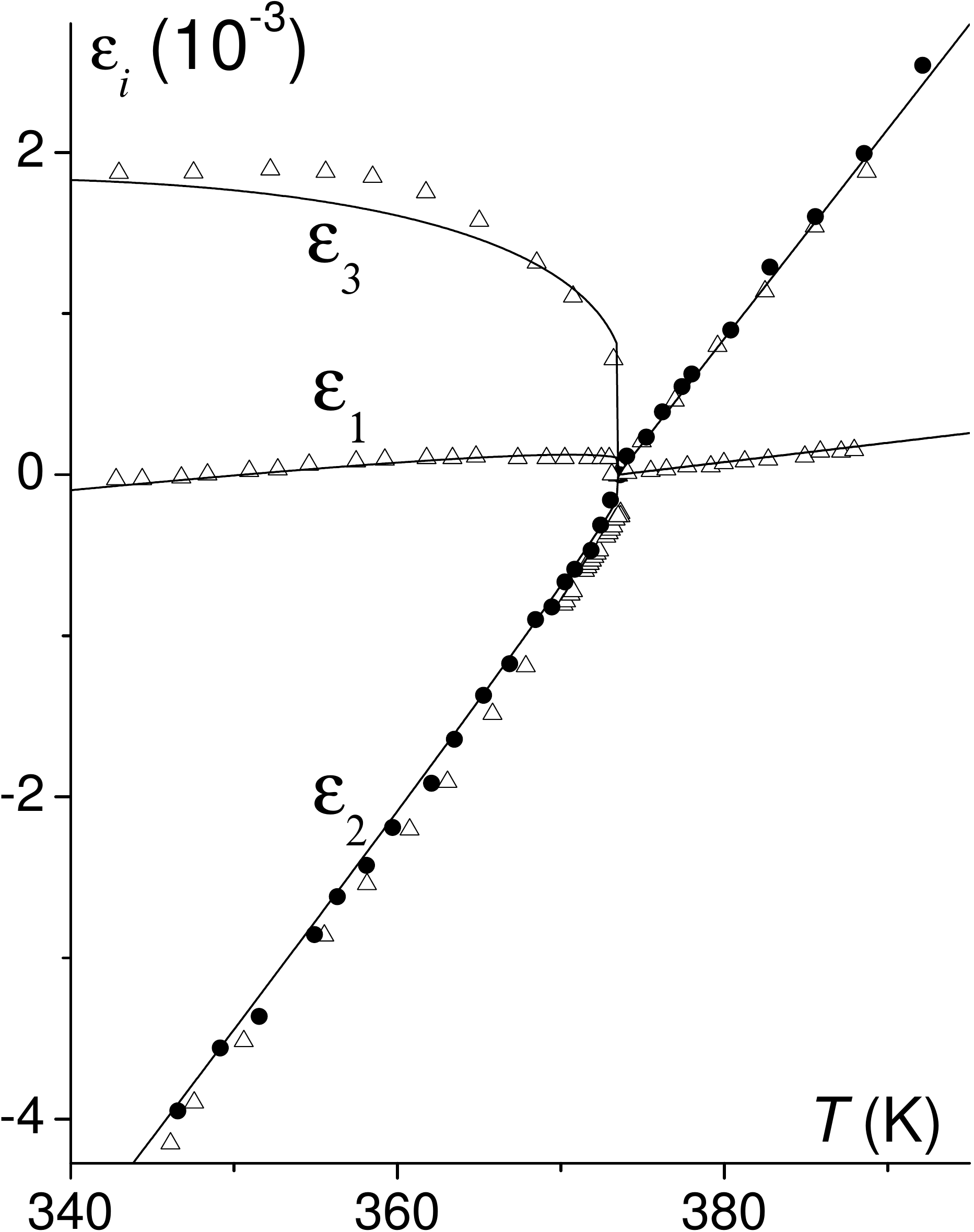}
		\hspace{5ex}
		\includegraphics[height=0.45\textwidth]{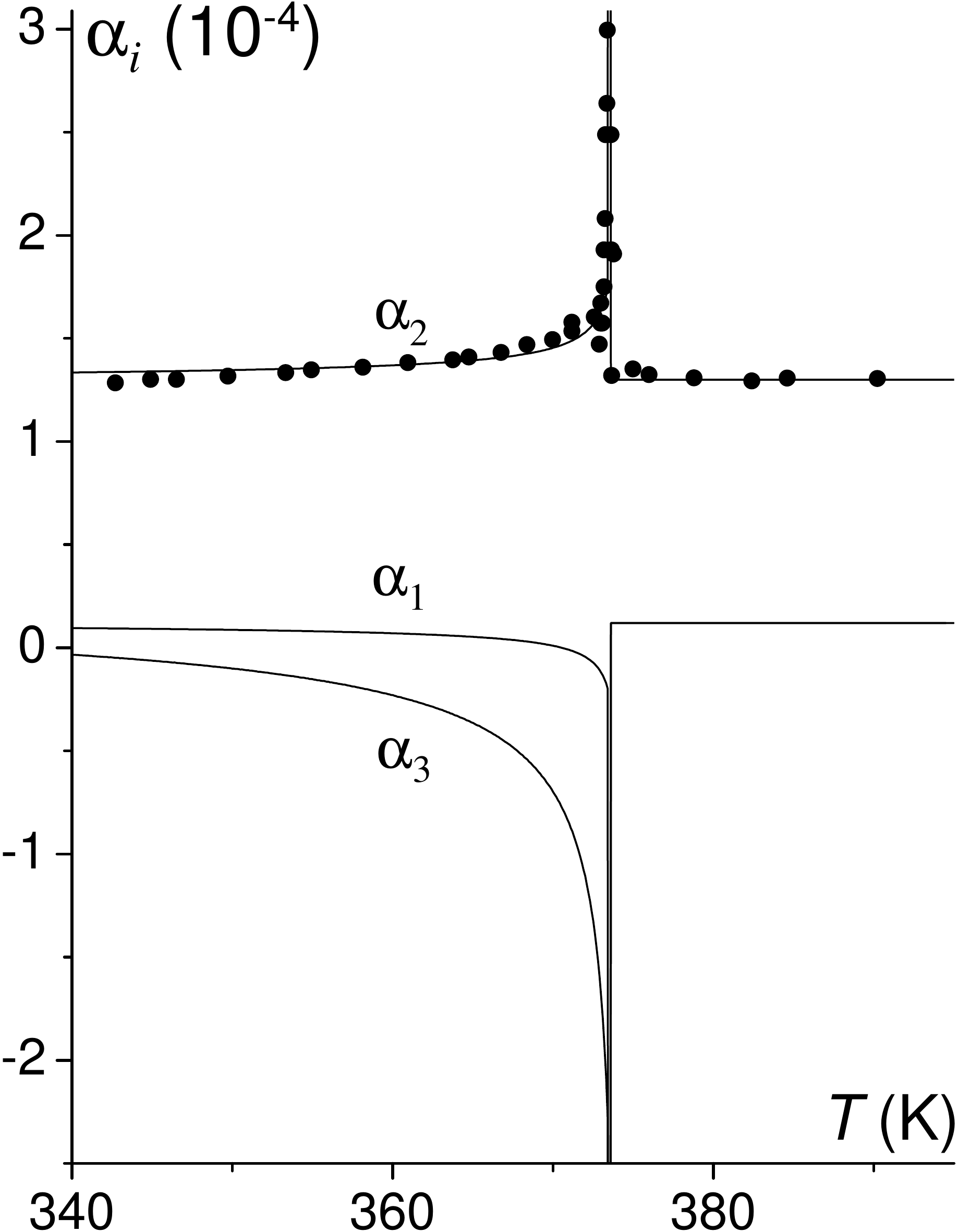}}
	\caption{Temperature dependences of the macroscopic lattice strains and of the corresponding thermal expansion coefficients of squaric acid at ambient pressure. Lines: the theory; symbols are experimental points taken from \cite{ehses:81} ($\bigtriangleup$) and \cite{johansen:84} ($\bullet$). } \label{fig-eps-thermalex}
\end{figure}

The temperature dependences of the diagonal lattice strains  $\eps_i$ and the corresponding thermal expansion coefficients are shown in figure~\ref{fig-eps-thermalex}. 
The coefficients were calculated by numerical differentiation of the strains with respect to temperature.

%They are obtained by extrapolation of the linear dependences of $\eps_i$ and $\mu'$ from the paraelectric phase down to the ordered phase and subtracting these extrapolated dependences from the entire curves. This procedure was performed both for theoretical and experimental data.

A clear anisotropy of the thermoelastic properties of squaric acid within the $ac$ plane and in the perpendicular direction is seen. At temperature lowering, the strain $\eps_2$ has a downward jump at the transition and a negative anomalous part in the ordered phase. The strains $\eps_1$ and $\eps_3$, on the other hand, have upward jumps and positive anomalous parts. As is shown above [see equation (\ref{eps-anom})], the anomalous contributions to the macroscopic strains are strictly proportional to the square of the order parameter~$\eta^2$. 
The thermal expansion coefficients $\alpha_1=\alpha_3$ in the paraelectric phase are by one order of magnitude smaller than $\alpha_2$; their anomalies at the transition point are of different signs.

\begin{figure}[!t]	
	\centerline{\includegraphics[height=0.3\textwidth]{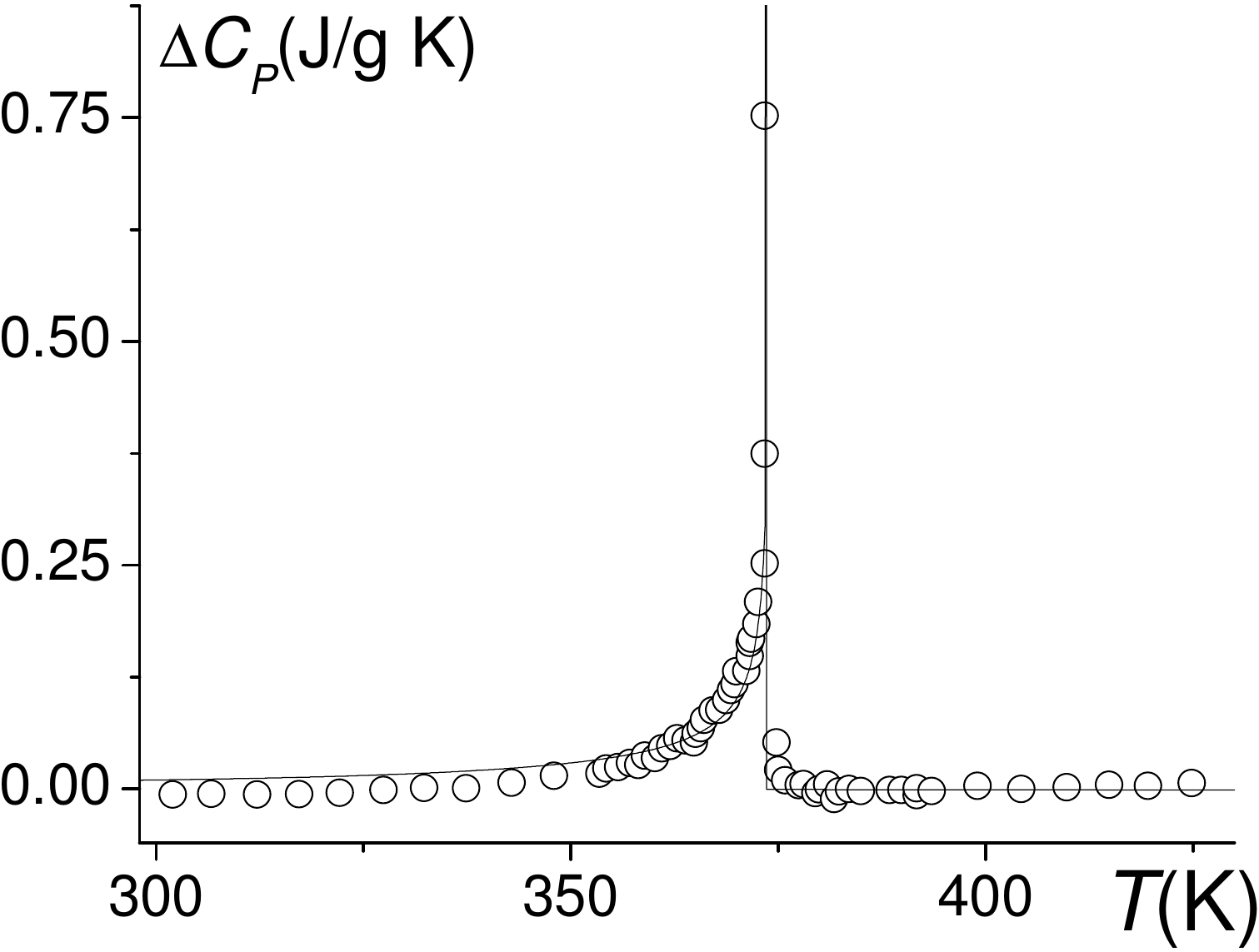}}
	\caption{Temperature dependence of the specific heat of squaric acid. Line: the theory; symbols are experimental points derived from the data of \cite{barth:79} as described in text. } \label{fig-Cp}
\end{figure}

The specific heat of the proton subsystem is calculated by numerical differentiation of the entropy (\ref{entropy}) with respect to temperature
\[
\Delta 
C_p=-\frac TM\left(\frac{\partial \Delta S}{\partial T}\right)_p,\]
where $M=114.06$~g/mol is the molar mass of squaric acid. The corresponding temperature curve is given in figure~\ref{fig-Cp}. The experimental points for the anomalous part of the specific heat are obtained by subtracting the
regular part, best described as a slightly non-linear curve $C_{\text{reg}}=-0.48803 + 0.00738 T - 7.3\cdot 10^{-6}T^2$ (J/g K), from the total specific heat as it was measured in \cite{barth:79}. One can see that a satisfactory agreement with experiment is obtained, even though the specific heat was not directly involved in the above described  fitting procedure. 

\begin{figure}[!b]
\centerline{\includegraphics[height=0.3	\textwidth]{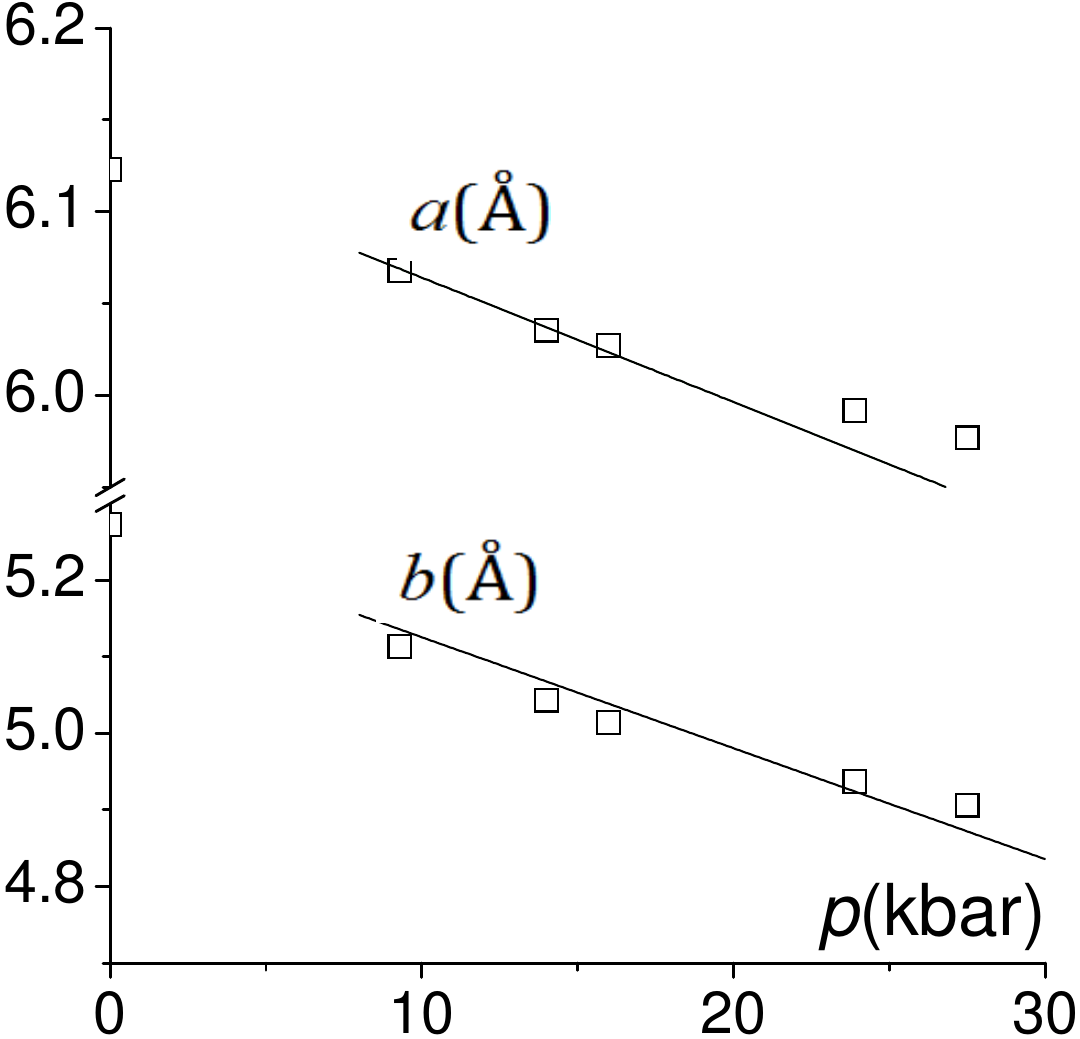}}
	\caption{Lattice constants at 292~K  as functions of hydrostatic pressure. Lines: the theory; symbols are experimental points taken from  \cite{katrusiak:86}.  } \label{fig-eps1eps2-hydro}
\end{figure}

The calculated hydrostatic pressure dependences of the paraelectric lattice constants are shown in figure~\ref{fig-eps1eps2-hydro}. The lattice constants were determined as $a=a_0[1+\eps(1)]$, $b=b_0[1+\eps(2)]$, where $a_0=6.137$~\AA, $b_0=5.327$~\AA\ are the values of the lattice constants just above the transition point at ambient pressure \cite{ehses:81}.  A good agreement with experiment is obtained. 

\begin{figure}[!t]
\centerline{\includegraphics[height=0.3\textwidth]{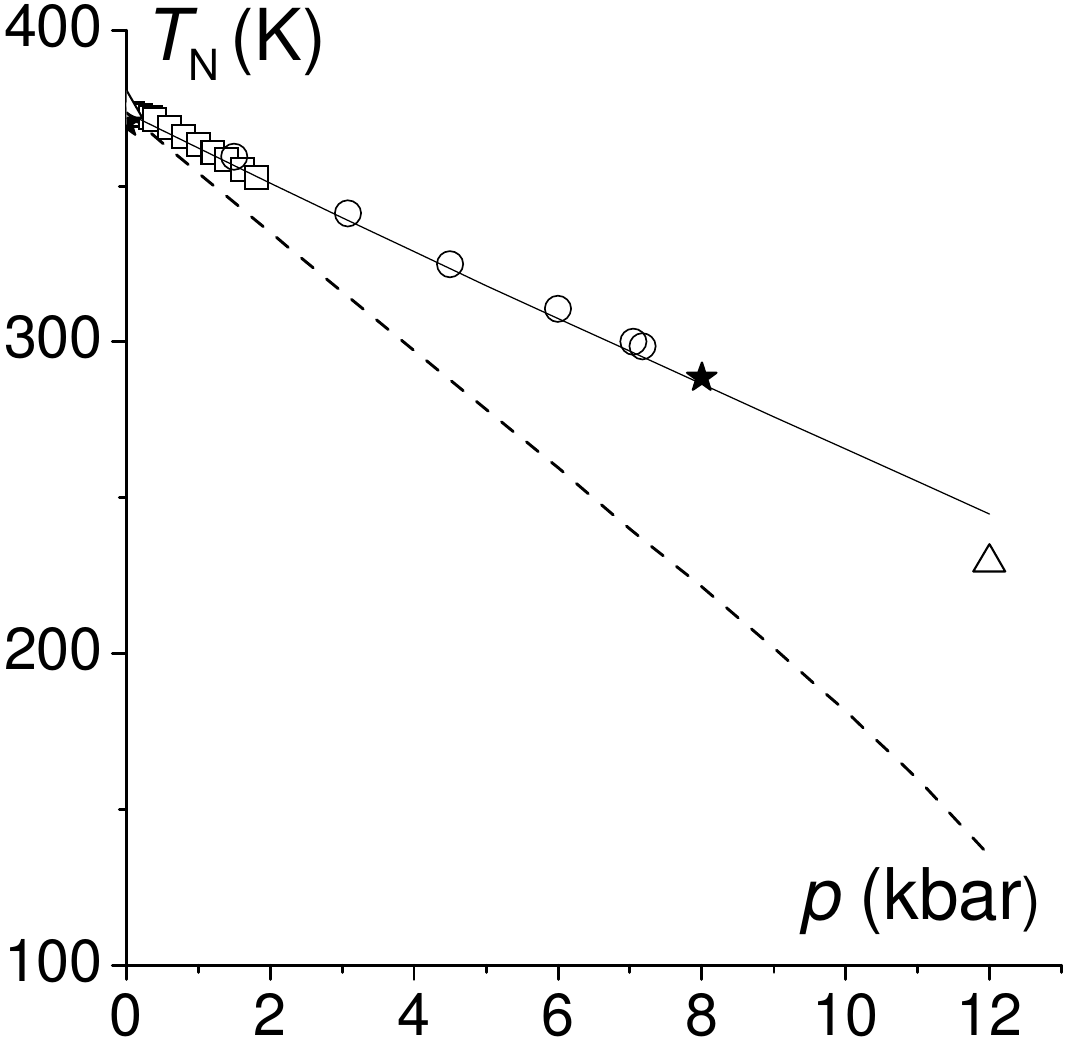}}
	\caption{Transition temperature of squaric acid as a function of hydrostatic pressure. Symbols are experimental points taken from  \cite{yasuda:78} ($\square$), \cite{samara:79} ($\circ$), \cite{moritomo:91} ($\bigtriangledown$) and \cite{mcmahon:90} ($\star$). The dashed and solid lines: the theory, calculated with equations (\ref{kdp-Slater}), (\ref{kdp-long}) and with equations (\ref{sqa-Slater}), (\ref{sqa-long}), respectively. } \label{fig-Tc-hydro}
\end{figure}

In figure~\ref{fig-Tc-hydro} we plot the hydrostatic pressure dependence of the phase transition temperature in squaric acid. As expected, the calculated transition temperature decreases with pressure (the dashed line). Quantitatively, however, completely non-satisfactory results are obtained. With the pressure variation of~$\mu'$ as observed experimentally \cite{mcmahon:90} and the parameters $\psi_i$ determined by fitting to the lattice strains below transition at ambient pressure, the rate, with which the calculated transition temperature decreases with hydrostatic pressure, $\partial T_c/\partial p=-19.5$~K/kbar, is nearly twice as large as the experimental one. The observed disagreement means, foremost, that equation~(\ref{kdp-long}) yields a too fast decrease of the long-range interaction parameter $\nu$, to which the theoretical values of the transition temperature are most sensitive. The pressure variation of the Slater-Takagi energies
is less important here.

Below we discuss a possible origin of the model inconsistency and ways to solve this problem. To this end, let us look closely at the obtained pressure dependence of $\nu$.

It is expected that the term $\sum_i\psi_i\eps_i$ in presence of high hydrostatic pressure would be positive, thereby leading to an increase of the long-range interaction parameter $\nu$ due to the reduction of the average interparticle distances in the 
compressed crystal. However, when the values of the parameters $\psi_i$ are chosen to fit to the experimental data \cite{ehses:81} for the anomalous spontaneous temperature behavior of the strains $\eps_i$ below the transition at ambient pressure, the sum $\sum_i\psi_i\eps_i$ in presence of high pressure becomes negative, not slowing, as expected, but enhancing the decrease of $\nu$ caused by the decrease of the H-site distance $\delta$. Of course, one cannot exclude a possibility that further measurements may yield the values of the strains $\eps_1(T)$, $\eps_3(T)$ different from those obtained in \cite{ehses:81} and unconfirmed yet by other measurements. The reasoning below, however, is based on the assumption that the data of \cite{ehses:81} are correct.

An obvious workaround but rather clumsy way to obtain the necessary pressure dependence of $\nu$  is to assume that there are some other high-pressure factors, not included into (\ref{kdp-Slater}) and (\ref{kdp-long}), and to include them empirically via the terms $k_{p}p$, namely
\begin{equation}
\label{sqa-Slater}
\eps=\eps_0\left[(1+\mu')^2+k_{p1}p\right], \quad w=w_0\left[(1+\mu')^2+k_{p1}p\right],
\end{equation}
and 
\begin{equation}
\label{sqa-long}
\nu=\nu_0\left[(1+\mu')^2+k_{p2}p\right]+\sum_{i=1}^3\psi_i\eps_i.
\end{equation}
We can speculate, for instance, that external pressure causes a redistribution of electron density, thereby changing the effective charges of the ions and, as a result, chanding the interactions between them. Introduction of two extra fitting parameters
$k_{p1}$ and $k_{p2}$ indeed allows us to describe the pressure variation of the transition temperature (see figure~\ref{fig-eps1eps2-hydro}, the solid line). At $k_{p1}=k_{p2}=0.0151$~kbar$^{-1}$ we obtain $\partial T_c/\partial p=-10.7$~K/kbar, in total agreement with experiment. Other combinations of $k_{p1}$ and $k_{p2}$ values can be found, also yielding the desired fit for the $T_c $ vs $p$ dependence. Accepting this, the already obtained good agreement with experiment for the system behavior at ambient pressure is not affected.

%A less speculative approach to the problem is, however, to recall that the essential non-linear temperature variation of the distance $\delta$ below the phase transition \cite{semmingsen:95} is neglected in the present model. If we expand it by considering $\delta$ as an independent thermodynamic variable, rather than a preset model parameter, then not only the theory will be more consistent and appropriate but also the above discussed  problem of the pressure variation of the long-range interaction parameter may be solved. Preliminary calculations show, however, that it does not suffice to simply determine $\delta$ (or, rather its displacement $\mu'$) by minimization of the thermodynamic potential, but the system Hamiltonian should be changed too. In particular, the terms like $c_\mu(\mu')^2$ and $\mu'\sum_i c_{i\mu}\eps_i$ should be included into $U_{\text{seed}}$ (here $c_\mu$ and $c_{i\mu}$ are components of the force-constant and internal-displacement tensors \cite{wu:95}).
%%The term $c_\mu(\mu')^2$ alone does not solve the problem, and the mixed terms %$\mu'\sum_i c_{i\mu}\eps_i$ are required. 
%This modification of the model is currently in progress and will be subject of a separate paper.

\section{Conclusions}
We present a modification to the proton ordering model, aimed at describing the effects associated with diagonal lattice strains in H-bonded antiferroelectric crystals of squaric acid. These effects include thermal expansion of the crystals, the appearance of spontaneous strain $\eps_1-\eps_3$ below the phase transition, and the shift of the transition temperature with hydrostatic pressure. 
Here, both the macroscopic lattice strains and the changes in the local geometry of hydrogen bonds are found to be essential. As usually, the quadratic dependence of the parameters of short-range and long-range interactions between protons on the H-site distance $\delta$ is assumed.

The deformational phenomena at ambient pressure are well described by the developed theory. On the other hand, the experimental dependence of the transition temperature on hydrostatic pressure can be described only if we assume that there are additional mechanisms to the pressure dependence of the interaction constants of the model, other than via the electrostriction interactions with the diagonal macroscopic strains and via the shortening of $\delta$, or if we suggest a further modification of the model, in which $\delta$ would be considered as an independent thermodynamic variable.

%
%  \lastpage
%  \end{document}
\newpage
\ukrainianpart

\title{Ефекти, пов'язані з діагональними деформаціями та геометрією водневих зв'язків, в антисегнетоелектричних кристалах квадратної кислоти}
\author{А.П. Моїна}
\address{
	Інститут фізики конденсованих систем НАН України, вул. Свєнціцького, 1,
	79011 Львів, Україна
}
%
%% якщо автор є один або автори є з однієї установи:
%
%  \author{1й Автор, 2й Автор, \ldots}
%  \address{Інститут\ldots}
%
%%

\makeukrtitle

\begin{abstract}
\tolerance=3000%
Запропоновано модифікацію моделі протонного впорядкування, яка має на свої меті опис фазового переходу та фізичних властивостей антисегнетоелектричних кристалів квадратної кислоти, що враховує вплив діагональних компонент тензора деформацій та локальної геометрії водневих зв'язків, а саме віддалі $\delta$ між  положеннями рівноваги протона на зв'язку. Теплове розширення, спонтанна деформація $\eps_1-\eps_3$ та теплоємність кристалу добре описуються запропонованою моделлю. Однак одночасний опис впливу гідростатичного тиску на температуру фазового переходу можливий лише при подальшому ускладненні моделі.%

\keywords
антисегнетоелектрик, водневий зв'язок, фазовий перехід, теплове розширення, гідростатичний тиск

\end{abstract}


\begin{thebibliography}{10}
\bibitem{semmingsen:95}
Semmingsen D., Tun Z., Nelmes R.J., McMullan R.K., Koetzle T.F., Z. Kristallogr., 1995, \textbf{210},\\ 934--947, \doi{10.1524/zkri.1995.210.12.934}.

\bibitem{semmingsen:77}
Semmingsen D., Hollander F.J., Koetzle T.F.,
J. Chem. Phys., 1977, \textbf{66}, 4405--4412, \doi{10.1063/1.433745}

\bibitem{hollander:77}
Hollander F.J., Semmingsen D., Koetzle T.F.,
J. Chem. Phys., 1977, \textbf{67}, 4825--4831. \doi{10.1063/1.434686}.

\bibitem{moritomo:91}  Moritomo Y.,  Tokura Y.,  Takahashi H.,  M\={o}ri  N.,  Phys. Rev. Lett., 1991, \textbf{67}, 2041-2044,\\ \doi{10.1103/PhysRevLett.67.2041}.

\bibitem{ehses:81}
Ehses K.H., Ferroelectrics, 1990, \textbf{108}, 277--282, \doi{10.1080/00150199008018770}.

\bibitem{klymachyov:97}
Klymachyov A.N.,  Dalal, N.S., Z. Phys. B: Condens. Matter, 1997, \textbf{104}, 651--656,
\doi{10.1007/s002570050503}.

\bibitem{samara:79}
Samara G.A., Semmingsen D., J. Chem. Phys., 1979, \textbf{71}, 1401--1407, \doi{/10.1063/1.438442}.

\bibitem{yasuda:78}  Yasuda N., Sumi K., Shimizu H.,  Fujimoto S.,  Okada K.,  Suzuki I.,  Sugie H.,  Yoshino K.,   Inuishi Y., J. Phys. C: Solid State Phys., 1978, \textbf{11}, L299--L303, \doi{10.1088/0022-3719/11/8/002}.


%\bibitem{moritomo:91}  Moritomo Y.,  Tokura Y.,  Takahashi H.,  M\={o}ri  N.,  Phys. Rev. Lett., 1991, \textbf{67}, 2041-2044,\\ \doi{10.1103/PhysRevLett.67.2041}.


\bibitem{matsushita:80}
Matsushita E., Matsubara T., Progr. Theor. Phys., 1980, \textbf{64}, No.~4,  1176--1192, \doi{10.1143/PTP.64.1176}.

\bibitem{matsushita:81}
Matsushita E., Matsubara T., Ferroelectrics, 1981, \textbf{39}, 1095--1098, \doi{10.1080/00150198108219573}.

\bibitem{matsushita:82}
Matsushita E., Matsubara T., Progr. Theor. Phys., 1982, \textbf{68}, No.~6,  1811--1826, \doi{10.1143/PTP.68.1811}.

\bibitem{chaudhuri:90}
Chaudhuri B.K.,  Dey P.K.,  Matsuo T.,
Phys. Rev. B, 1990, \textbf{41}, 2479--2489, \doi{10.1103/PhysRevB.41.2479}.

\bibitem{maier:82}
Maier H.-D.,  M\" user H.E., Petersson J.,  Z. Phys. B: Condens. Matter, 1982, \textbf{46}, 251--260,\\ \doi{10.1007/BF01360302}.


\bibitem{deininghaus:81}
Deininghaus U., Mehring M., Solid State Commun.,  1981, \textbf{39}, No.~11, 1257--1260,\\ \doi{10.1016/0038-1098(81)91126-1}


\bibitem{ishizuka:11}
Ishizuka H.,  Motome Y.,  Furukawa N.,  Suzuki S.,
Phys. Rev. B, 2011, \textbf{84}, 064120 (6 pages),\\ \doi{10.1103/PhysRevB.84.064120}.

\bibitem{vijigiri:20}
Vijigiri V.,  Mandal S., 
J. Phys.: Condens. Matter, 2020, \textbf{32}, 285802, \doi{10.1088/1361-648X/ab7ba1}.


\bibitem{mcmahon:90}
Mcmahon M.I.,  Piltz R.O., Nelmes R.J., Ferroelectrics, 1990, \textbf{108},
277--282, \doi{10.1080/00150199008018770}.


\bibitem{stasyuk:99} Stasyuk I.V., Levitskii R.R.,  Moina A.P., Phys. Rev. B, 1999, \textbf{59}, 
8530--8540, \doi{10.1103/PhysRevB.59.8530}.


\bibitem{moina:11}
Moina A.P., Levitskii R.R., Zachek I.R., Condens. Matter Phys., 2011, \textbf{14},  43602 (18 pages), \\\doi{10.5488/CMP.14.43602}.


\bibitem{blinc:87}
Blinc R.,  Zeks B., Ferroelectrics, 1987,  \textbf{72}, 193--227, \doi{10.1080/00150198708017947}. 


\bibitem{semmingsen:74}
Semmingsen D., Feder J., Solid State Commun., 1974, \textbf{15}, 1369--1372,
\doi{10.1016/0038-1098(74)91382-9}.

\bibitem{moritomo:90}
Moritomo Y., Koshihara S., Tokura Y., J. Chem. Phys., 1990, \textbf{93}, 5429--5435,
\doi{10.1063/1.459639}.

\bibitem{stasyuk:00}
Stasyuk I.V., Levitskii R.R., Zachek I.R., Moina A.P.,  Phys. Rev. B, 2000, \textbf{62}, 6198--6207, \\ \doi{10.1103/PhysRevB.62.6198}.

\bibitem{stasyuk:01}
Stasyuk I.V., Levitskii R.R., Moina A.P., Lisnii B.M., Ferroelectrics, 2001, \textbf{254}, 213--227, \\ \doi{10.1080/00150190108215002}.

\bibitem{blinc:66}
Blinc R.,  Svetina S., Phys. Rev., 1966, \textbf{147}, 430--438, \doi{10.1103/PhysRev.147.430}.




\bibitem{levitskii:04}
Levitskii R.R.,  Lisnii B.M., Phys. Status Solidi B, 2004, \textbf{241}, 1350--1368, \doi{10.1002/pssb.200301995}.


\bibitem{torstveit} Torstveit S.,
Phys. Rev. B, 1979, \textbf{20}, 4431--4441, \doi{10.1103/PhysRevB.20.4431}.

\bibitem{rehwald:78}
Rehwald W., Vonlanthen A., Phys. Status Solidi B, 1978, \textbf{90},  61--66
\doi{10.1002/pssb.2220900106}. 

\bibitem{yamanaka:87}
Yamanaka A., Tatsuzaki I., J. Phys. Soc. Jpn., 1987, \textbf{56},  1043--1050, \doi{10.1143/JPSJ.56.1043}.



\bibitem{katrusiak:86} Katrusiak A.,  Nelmes R.J., J. Phys. C: Solid State Phys., 1986, \textbf{19}, L765--L772, \\ 
\doi{10.1088/0022-3719/19/32/001}.

\bibitem{johansen:84}
Johansen T. H., Feder J., J\o ssang T., 
Z. Phys. B: Condens. Matter, 1984, \textbf{56}, 41--49, 
\doi{10.1007/BF01470211}. 

\bibitem{mehring:81} 
 Mehring M., Becker J.D., Phys. Rev. Lett., 1981, \textbf{47}, 366--370, 
 \doi{10.1103/PhysRevLett.47.366}.

\bibitem{barth:79}
 Barth E.,  Helwig J.,  Maier H.-D.,  M\"{u}ser H.E.,  Petersson J.,
 Z. Phys. B: Condens. Matter, 1979, \textbf{34}, 393--397, 
  \doi{10.1007/BF01325204}.

%\bibitem{wu:95}
%Wu X.,   Vanderbilt D.,  Hamann D.R., Phys. Rev. B, 2005, \textbf{72}, 035105 (13 pages),\\
%\doi{10.1103/PhysRevB.72.035105}.




\end{thebibliography}
\end{document}